\documentclass[pdflatex,sn-mathphys-num]{sn-jnl}

\usepackage{amsmath, amssymb}
\usepackage{graphicx}
\usepackage{hyperref}
\usepackage[numbers]{natbib}
\usepackage[utf8]{inputenc}
\usepackage{verbatim}
\usepackage{graphicx}
\usepackage{booktabs}
\usepackage{tabularx}
\usepackage{subcaption}
\usepackage[svgnames]{xcolor}
\usepackage{ulem}

\newcommand{\rev}[1]{{\color{black}#1}}

\newcommand{\bidc}[1]{\mathrm{BiDC}_{#1}}
\newcommand{\monodc}[1]{\mathrm{MonoDC}_{#1}}

\title{\rev{Leveraging Content Producer Networks and User Perception to Detect Online Discursive Communities}}

\author[1,*]{Stefano Guarino}
\equalcont{These authors contributed equally to this work.}
\author[2]{Ayoub Mounim}
\author[3,4,5]{Guido Caldarelli}
\author[6,1,7]{Fabio Saracco}
\equalcont{These authors contributed equally to this work.}

\affil[1]{National Research Council of Italy, Institute for Applied Mathematics, Rome, Italy}
\affil[2]{LUISS ``Guido Carli'' University, Data Lab, Rome, Italy}
\affil[3]{National Research Council of Italy, Institute of Complex Systems, Rome, Italy}
\affil[4]{Ca’ Foscari University of Venice, Venice, Italy}
\affil[5]{London Institute for Mathematical Sciences, London, United Kingdom}
\affil[6]{``Enrico Fermi'' Research Center, Rome, Italy}
\affil[7]{IMT School for Advanced Studies Lucca, Lucca, Italy}

\affil[*]{Corresponding author: \texttt{stefano.guarino@cnr.it}}

\date{}

\begin{document}

\maketitle

\rev{
\begin{abstract}

Online discussions are often characterized by strong behavioral asymmetries: a relatively small fraction of users actively produces content, while the majority primarily consumes and redistributes it.
Here we propose a community-detection framework for online social networks that exploits this asymmetry by first identifying and clustering a set of leading users, and then extending the resulting labels to the broader user base.
We introduce two complementary strategies to cluster leaders, one based on their mutual interactions and the other on audience overlap, both relying on entropy-based filtering to separate signal from noise.
We evaluate the framework on three major Italian political debates on Twitter/X, using public figures--identified through the pre-2022 verification system--as leaders, and known affiliations of political actors as ground truth labels.
Compared with standard baselines, the proposed approach yields more coherent and interpretable communities aligned with political structures, with the two variants respectively recovering parties and coalitions.
Activity-based criteria for selecting leaders produce qualitatively similar but consistently weaker results, particularly at the coalition level.
Overall, our findings show that creating statistically validated networks of publicly recognized figures, whose off-platform roles constrain and stabilize their online behavior, provide a strong basis to identify discursive communities on social media.
Although developed for Twitter/X, the approach is conceptually general, as it leverages structural asymmetries common to many online platforms.

\end{abstract}
}

\section*{Introduction}

Online social platforms have become crucial arenas for public discourse, with communities forming organically around shared narratives, values, and ideological affinities~\citep{Conover2011a,Conover2011b,Conover2012}. 
While these discursive communities can offer solidarity and foster public engagement~\citep{GonzalezBailon2013, Andalibi2018, Morini2025}, 
they also risk becoming echo chambers—insulated environments that reinforce existing beliefs and suppress dissenting perspectives~\citep{jamieson08echo,Garrett2009,delvicario2016,Zollo2017Debunking,Villa2021,Morini2021,Pratelli2024}. 
Understanding the structural dynamics that govern these communities is therefore essential to evaluating how information circulates and how opinions crystallize in digital spaces.

Traditional approaches to community detection in social networks have evolved significantly, incorporating network topology, account metadata, and behavioral signals.
For instance, some methods enhance classical algorithms by integrating account characteristics~\citep{citraro2019eva}, while others acknowledge that platforms like Twitter/X function less as egalitarian many-to-many networks, and more as hierarchies dominated by a small number of highly visible content creators~\cite{Watts2007,Wu2011,Hilbert2017}. Within this structure, influence is disproportionately concentrated, making the identification of these “elite” accounts central to the analysis of online discourse~\cite{Becatti2019d,Caldarelli2020b,Radicioni2021a}.

\rev{
Building on this body of work, the first contribution of this article is the definition of a general framework for detecting online discursive communities, formalizing the idea that online debates are driven by a relatively small set of users who actively produce content.
Because content production entails greater exposure and commitment—especially in political discussions—these users tend to adopt more consistent and publicly interpretable positions.
Once content producers are identified, the affiliation of less active users can be inferred from their interaction patterns, even when individual behavior is sparse.

The proposed framework operationalizes this rationale in a relatively simple way.
We partition the user set $U$ into two disjoint subsets, $U=\top\cup\bot$, cluster the content creators (the set $\top$) by applying a standard community-detection algorithm to an appropriately defined network $G=(\top,E)$, and then extend the resulting labels to users in $\bot$ based on their interactions with $\top$.
}

\rev{
A key modeling choice, however, concerns the construction of the edge set $E$, which defines how relationships among content creators are represented.
We consider two alternatives: one based on statistically significant interactions directly observed among users in $\top$, and one based on a validated projection of the bipartite network linking content creators in $\top$ to their audiences in $\bot$.
These two variants define \textit{Monopartite-network-induced Discursive Communities} ($\monodc{\top}$) and \textit{Bipartite-network-induced Discursive Communities} ($\bidc{\top}$), respectively.}
Both methods rely on maximum-entropy null models~\citep{Cimini2019} to statistically validate inferred connections and to filter out noise--a well-known challenge in the analysis of online social networks~\citep{Becatti2019d,Declerck2022a}.
\rev{
Statistical filtering plays a limited role in $\monodc{\top}$, but is essential in $\bidc{\top}$ to remove spurious correlations induced by heterogeneous audience activity.
}

\rev{Beyond the construction of $E$, the framework admits different criteria to identify the set $\top$ of content creators.
From this perspective, the second contribution of this article is less methodological and more substantive: we show that, although social-media relevance can be acquired in an anonymous and platform-driven manner, ideological communities in political debates remain predominantly anchored to publicly recognizable political figures.

Accordingly, we focus on Twitter's verified accounts ($\top=V$), which, at the time covered by our datasets, identified public figures whose activity on the platform was closely tied to their off-platform roles.
All datasets analyzed precede the introduction of paid verification in November 2022, and therefore correspond to this original, institutionally grounded meaning of verification~\cite{Haman2023}.

We compare this choice with two activity-based criteria derived from the data.
We consider \textit{influential} users ($\top=I$) as defined by~\citep{GonzalezBailon2013}, based on follower structure and mention activity, and \textit{high-$h$-index} users ($\top=H$), obtained by adapting the Hirsch index~\citep{Hirsch2005} to retweet dynamics.
These data-driven selections serve as baselines to evaluate the added value of anchoring the analysis on publicly recognized figures.

We analyze three datasets covering major political events in Italy in 2022, where discursive communities are expected to mirror political alignments.
When evaluated against the known affiliations of clearly identifiable political actors, $\monodc{V}$ captures party-level structures, whereas $\bidc{V}$ is more effective at identifying coalitional configurations.
Standard community-detection algorithms applied to the complete retweet network yield noisier and less interpretable partitions, while activity-based criteria perform better than these baselines but remain less reliable than anchoring the analysis on verified accounts, especially at the coalition level.

We argue that this is not an \textit{effect} of verification, but, again, a consequence of the fact that, historically, only publicly recognizable figures were ``verifiable'', a requisite that in turn constrained and stabilized their online behavior.
In this sense, our results show that, even in environments where visibility can be gained in largely anonymous and platform-driven ways, ideological communities in political debates remain primarily structured around actors endowed with public recognizability beyond the platform.
By foregrounding this mechanism, we propose and analyze a framework that, while developed for Twitter/X, is conceptually general, as it leverages asymmetries between content producers and audiences that characterize many contemporary online platforms.

Finally, we show that a carefully selected subset of users and interactions captures a large fraction of the information needed to reconstruct political communities, offering practical insights for designing efficient data-collection strategies under increasingly restrictive platform access conditions.}

\rev{
\subsubsection*{Related work}
A substantial body of research has investigated the structure of the online discourse on Twitter and other social media platforms by analyzing retweet networks, information cascades, and statistically validated interaction patterns (e.g.,~\cite{garimella2018quantifying,Becatti2019d,cinelli2021echo,lorenzspreen2023digitalmedia}). 
These studies have consistently shown, in particular, that political debates on social media exhibit strong polarization and modular structure, often aligned with ideological or partisan divides.

A central methodological challenge in this literature concerns the high level of noise and heterogeneity in user activity.
To address this issue, several works have adopted statistical filtering techniques based on null models to extract significant interaction patterns from bipartite or projected networks.
In particular, maximum-entropy null models have been widely used to preserve local constraints—such as node degrees—while filtering out correlations that can be explained by random activity alone~\cite{Cimini2019,Saracco2017}.
These approaches have been shown to uncover meaningful mesoscale structures in retweet, mention, and user-content networks, enabling the identification of latent ideological alignment from interaction data~\cite{Becatti2019d,Caldarelli2020b,Radicioni2021a,Mattei2022,Pratelli2024}.

Within this line of work, the Italian political context has received special attention.
Becatti et al.~\cite{Becatti2019d} have shown that entropy-based filtering improves the interpretability and stability of political communities compared to unfiltered network representations (the method proposed therein is called $\bidc{V}$ in the present manuscript).
Related studies have further demonstrated that statistically validated user--hashtag bipartite projections can reveal well-defined discursive communities during national electoral campaigns, even in the absence of explicit ideological labels~\cite{Radicioni2021a}.

A parallel strand of research has focused on the role of \textit{influential} or \textit{central} users in shaping the online discourse.
Early work has defined influence primarily in terms of activity-based or structural measures, such as the number of followers, retweets, mentions, or centrality in interaction networks~\cite{Watts2007,Hilbert2017, Wu2011}.
Other studies have explored the diffusion power of opinion leaders and elites in online debates, often highlighting the disproportionate impact of a small fraction of highly visible users~\cite{bakshy2011everyone}.

Despite these advances, two important gaps persist in the literature.
First, most community-detection approaches implicitly treat users symmetrically, without explicitly accounting for the strong asymmetry between content producers and content consumers that characterizes participation on social media--in general, and especially for political debates.
Second, benchmarks for evaluating communities are often based on internal network metrics, such as modularity or stability, rather than on externally validated ground truth labels.
% of political affiliation.
As a result, it remains unclear (i) whether institutional signals of authority--such as pre-2022 verified status on Twitter--provide particularly informative anchors for community detection, and (ii) how alternative criteria for identifying leading users affect the resulting mappings when embedded into statistically validated pipelines.
The present work addresses these gaps by proposing a leader-based framework that explicitly separates content creators from their audiences and systematically compares different strategies for selecting politically relevant producers of content.
}

\section*{Methods}

We investigate how online discursive communities form around publicly relevant users on Twitter/X by analyzing large-scale interaction data.
While the framework we propose is general, we focus empirically on three major political events in Italy during 2022, a domain in which online debates tend to be especially structured and externally grounded.
Our approach builds on the observation that social-media debates are typically characterized by strong asymmetries in user behavior: a relatively small subset of users actively produces salient content, whereas the majority primarily consumes and redistributes it.
This asymmetry is particularly pronounced in political discussions, where public figures and institutional actors play a central role.
To exploit this structure, we adopt a leader-based framework in which community detection is first performed on a selected set of content creators and subsequently extended to the broader user base.

\rev{
\subsection*{Detection of political discursive communities}
}

To identify meaningful structures in online political debates, it is necessary to distinguish statistically significant interaction patterns from correlations induced by heterogeneous user activity and content popularity.
To this end, our approach relies on maximum-entropy null models as a principled statistical baseline for validating inferred network connections before applying community-detection algorithms.

Within this framework, an ensemble of random networks is defined by maximizing the Shannon entropy subject to a set of local constraints that reproduce selected properties of the observed data, such as degree sequences or activity levels~\citep{Jaynes1957,park2004statistical,Cimini2019}.
Empirical links are evaluated against their expected occurrence under the null model, and only statistically significant deviations are retained.
Model parameters are estimated by maximizing the likelihood of the observed network~\citep{Garlaschelli2008}, and statistical significance is assessed using false discovery rate correction~\citep{Benjamini1995}.

This maximum-entropy approach is grounded in information theory and statistical mechanics and has proven effective across a wide range of networked systems~\citep{Becatti2019d,Caldarelli2020b,Caldarelli2021,Radicioni2021a,Radicioni2021b,Mattei2021,Guarino2021,Mattei2022,Bruno2022,Declerck2022a,Declerck2022b,Pratelli2024}.
In the following, we introduce two complementary methods that instantiate this framework in different ways, depending on how interactions among selected content creators are inferred.
\rev{
Further methodological details are reported in Section~\ref{app:maxent} of the Appendix.
}

In the analyses presented here, we focus on verified accounts as a concrete instantiation of content creators, as--prior to the introduction of paid verification--they largely corresponded to publicly recognized figures.
However, the methods described below are general and can be applied to any set $\top$ of users identified as debate leaders.

\rev{
\subsubsection*{Construction of the $\monodc{V}$ network}
}

\begin{figure}[htbp!]
    \centering
    \includegraphics[width=\textwidth]{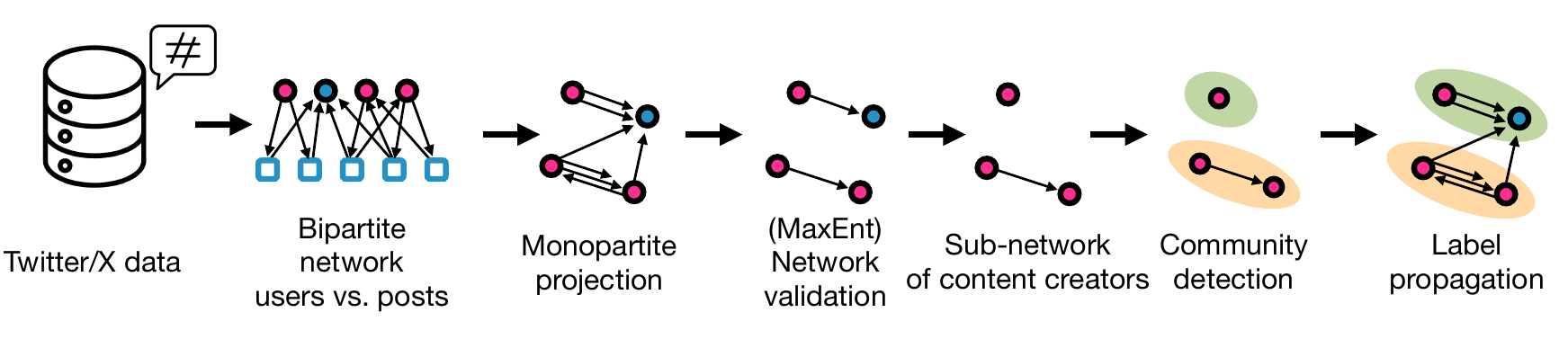}
    \caption{\textbf{$\monodc{}$'s pipeline.} Starting from the data from Twitter/X, we build a bipartite directed network between accounts and posts; a directed arrow from a user \rev{(a circle)} to a post \rev{(a square)} represents authorship, while in the opposite direction represents a retweet. Then, the content of the bipartite network is projected into a monopartite directed network of users. Such a network is further statistically validated using the expectations of a maximum entropy null model for bipartite directed networks. Then, the sub-network relative to \rev{the selected content creators (magenta circles)} is analysed, and communities are detected using standard algorithms. Finally, the so-obtained labels are propagated to standard users \rev{(blue circles)} through a label propagation algorithm. \rev{The icon representing Twitter/X data is obtained by superimposing the icons ``Data Base \#7994662'' by Farrih Icon and ``Hashtag \#4315467'' by Alex Burte, both from Noun Projects.}}
    \label{fig:monodc_cartoon}
\end{figure}

The first method, $\monodc{V}$ (Monopartite-network-induced Discursive Communities; Fig.~\ref{fig:monodc_cartoon}), focuses on retweet interactions among verified users.
We start by constructing a bipartite directed network between users and tweets, where directed edges represent authorship and retweet actions.
\rev{
At this stage, we consider all verified users together with all users who retweeted at least one verified account in the dataset under study.
}
This structure is then projected onto a weighted directed user-user network, in which a link from user $i$ to user $j$ encodes the number of times $i$ retweeted content authored by $j$.

To separate meaningful interaction patterns from noise--an intrinsic feature of online social activity--we apply a maximum-entropy null model tailored to bipartite directed networks.
The model uses the observed posting and retweeting activity of users as constraints while randomizing all other aspects, providing a principled statistical baseline.
\rev{
For each pair of users \(i,j\), we compute the probability of the observed retweeting weight under the null model, and retain the edge only if its p-value is statistically significant (more details are present in Section \ref{app:maxent} of the Appendix).
}

Community detection is then performed on the subnetwork of the validated network induced by verified users.
The presented results have been obtained with a modularity-based algorithm, but we verified that alternative algorithms yield essentially the same results.
Finally, the detected community labels are propagated to non-verified users using a standard label-propagation scheme~\citep{Raghavan2007a}.

\rev{
\subsubsection*{Construction of the BiDC\(_V\) network}
}

\begin{figure}[htbp!]
    \centering
    \includegraphics[width=\textwidth]{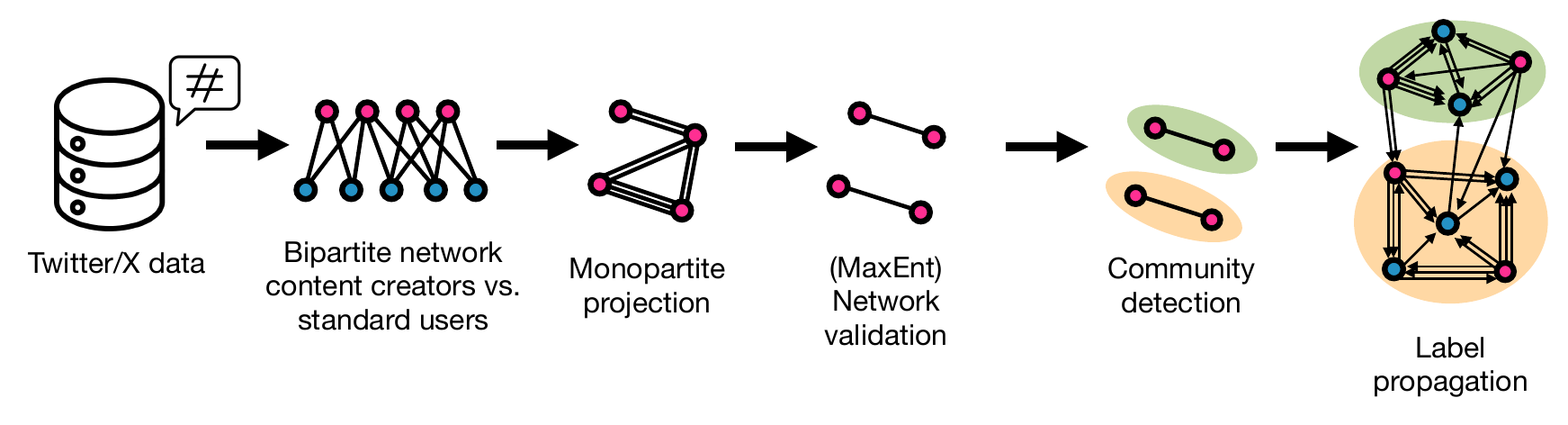}
    \caption{\textbf{$\bidc{}$'s pipeline.} Starting from the data from Twitter/X, we build a bipartite undirected network between the accounts of the \rev{content creators} and the one of standard users; a link is present if the standard users retweeted the given \rev{content creator} at least once. Then, the information of the bipartite network is projected into a monopartite network of \rev{content creators}. Such a network is further statistically validated using the expectations of a maximum entropy null model for bipartite undirected networks. Finally, the so-obtained labels are propagated to standard users through a label propagation algorithm. The procedure was first proposed in~\cite{Becatti2019d}. \rev{The icon representing Twitter/X data is obtained by superimposing the icons ``Data Base \#7994662'' by Farrih Icon and ``Hashtag \#4315467'' by Alex Burte, both from Noun Projects.}}
    \label{fig:bidc_cartoon}
\end{figure}

The second method, $\bidc{V}$ (Bipartite-network-induced Discursive Communities; Fig.~\ref{fig:bidc_cartoon},~\cite{Becatti2019d}), infers relationships among verified users indirectly, based on the similarity of their audiences.
Here, we construct a bipartite undirected network connecting verified users to all users who retweeted them at least once during the event.

This bipartite structure preserves the heterogeneity between content creators and their audiences.
To extract meaningful similarities between verified users, we apply a maximum-entropy null model for bipartite undirected networks, which accounts for differences in audience sizes and engagement patterns.
Statistically significant similarities are retained, yielding a validated projection onto a network of verified users.

Community detection is performed on this projected network, and the resulting labels are propagated to non-verified users according to the composition of their retweet activity.
As in $\monodc{V}$, the propagation step allows us to assign community memberships to the entire user set starting from a limited number of leader nodes.\\

\rev{
\subsection*{Evaluation}
}

We evaluate the quality of the detected communities using the V-measure ($\mathrm{VM}_{\beta}$)~\cite{rosenberg2007v}, an information-theoretic metric that quantifies the agreement between two partitions based on their mutual information.
The V-measure is well suited to our setting, as it remains interpretable across different levels of partition granularity and explicitly balances \textit{homogeneity}--how well each cluster contains only members of a single class--and \textit{completeness}--whether all members of a given class are assigned to the same cluster--through the parameter $\beta$.
For $\beta=1$, $\mathrm{VM}_{1}$ reduces to the Variation of Information~\citep{Meila2007}, normalized by the total entropy of the partitions.
By varying $\beta$ over $[0,+\infty)$, we account for both finer-grained and more aggregated political structures, such as political movements/trends within parties or political coalitions. \rev{More details can be found in Section \ref{app:v_measure} of the Appendix.}

As a baseline, we also apply standard community-detection algorithms directly to the complete retweet network.
\rev{
This network is defined as a directed, weighted graph including all active users, where edges represent the number of retweets between pairs of users.
Unlike $\monodc{V}$ and $\bidc{V}$, this representation is not statistically filtered and serves to isolate the contribution of entropy-based validation and leader selection.
}

\rev{
Although MonoDC\(_V\) and BiDC\(_V\) ultimately rely on Louvain for final partitioning, the comparison with Louvain applied to the complete retweet network serves a different purpose. 
In our frameworks, Louvain operates \textit{after} statistically validated filtering, and \textit{with} an informative initialization provided by verified seeds. 
Running Louvain directly on the raw network therefore represents a baseline that isolates the contribution of our preprocessing and seed selection. 
This allows us to quantify the extent to which discursive communities emerge from the raw interaction structure alone versus from statistically validated and leader-anchored representations.}\\

\subsection*{Data}

We analyze three datasets covering the Italian presidential election, a government crisis, and the 2022 general elections.
Each dataset consists of millions of tweets and hundreds of thousands of users, collected via Twitter’s Academic \rev{Application Programming Interface (API)} using event-specific keywords; \rev{more details about the queries used for the data collection can be found in Section \ref{app:data_collection} of the Appendix.}.
Table~\ref{tab:data} summarizes the main characteristics of the datasets.

\rev{
Among the verified accounts appearing in each dataset, we manually annotated those corresponding to politically active public figures.
}
Annotations were performed at both the party and coalition level and provide an external ground truth against which we assess the performance of our methods.

\rev{
\subsubsection*{Manual annotation of verified accounts}

For each event, we manually annotated the political affiliation of verified users to construct the ground truth used in our evaluation.
Only verified accounts of political parties and politicians were annotated; accounts belonging to journalists, media outlets, institutions, or entertainers were excluded.
Each annotated user was assigned both a \textit{party label} and a \textit{coalition label}.
The party label was determined based on publicly available information from the user's Twitter profile and Wikipedia page, or from institutional sources such as party websites, parliamentary records and candidate lists.
The coalition label was assigned according to the official coalitions recognized during the 2022 Italian electoral cycle. 

Annotations were performed by one of the authors, while the remaining authors conducted random spot checks to assess consistency and correctness.
Verified accounts with no clear political alignment were included in the construction of both $\monodc{V}$ and $\bidc{V}$ networks, but were excluded from the ground truth and used only as unlabelled nodes in the evaluation.
}

\begin{table}[htbp]
    \centering
    \caption{A summary of the datasets considered in the present paper}
    \label{tab:data}
    \begin{tabularx}{\textwidth}{XXXXXXX}
    \toprule
    \textbf{Dataset} & 
    \textbf{Unique users} &
    \textbf{Unique tweets} &
    \textbf{Annotated users} &
    \textbf{Verified users} &
    \textbf{Influential users} &
    \textbf{\mbox{H-index$\geq$3} users}\\
    \midrule
    %%%%%%%%%%%%%%%%%%%%%%%%%
    \textbf{President}\tnote{1} &
    119.018 &
    711.497 &
    292 &
    1148 &
    5804 &
    1296\\
    %%%%%%%%%%%%%%%%%%%%%%%%%
    \textbf{Crisis}\tnote{2} &
    113.876 &
    1.414.181 &
    229 &
    1004 &
    7038 &
    2554\\
    %%%%%%%%%%%%%%%%%%%%%%%%%
    \textbf{Elections}\tnote{3} &
    222.356 &
    2.784.951 &
    239 &
    1555 &
    12184 &
    4382\\
    \bottomrule
    \end{tabularx}
    \begin{tablenotes}
    \item[1] Debate around the election of the President of the Republic. The political debate ended with the re-election of Sergio Mattarella on January 29, 2022.
    \item[2] Debate around the political crisis that led to the fall of the Draghi government on July 21, 2022.
    \item[3] Debate around the Italian general elections held on September 25, 2022, that led to Giorgia Meloni becoming the new Prime Minister.
    \end{tablenotes}
\end{table}

\rev{
\subsubsection*{Code and data availability}

Preprocessed data and code used for the analysis are available on GitLab at \url{https://gitlab.com/s.guarino/dico_analysis}.
}

\section*{Results}

\rev{
\subsection*{Verified accounts provide a strong political signal under the pre-2022 verification regime}
}

To assess the effectiveness of our framework, we focused on three political events that sparked intense debate on Italian Twitter in 2022: the re-election of the President of the Republic, the resignation of Prime Minister Draghi, and the general elections in September. 
Crucially, these datasets were compiled before the platform introduced paid verification, ensuring that verified accounts retained their original role as signals of institutional recognition rather than self-declared status.

We manually annotated a subset of verified accounts (between 229 and 292 per event) based on their publicly identifiable political affiliations.
When restricting the retweet networks to\rev{, first, the Greatest Connected Component (GCC), and then to} annotated users and evaluating the partition induced by their known party and coalition labels, we observed highly modular structures.
Modularity values exceeded 0.6 in all cases (cfr. Table~\ref{tab:bench}), indicating that retweet interactions among verified users are strongly structured along political lines.

\rev{
These results show that, even without any statistical validation, retweet interactions among the annotated verified users already separate clearly along party and coalition lines, indicating that well-defined political communities exist within this subset.
}

\begin{table}[h!]
    \centering
    \caption{Retweet networks of annotated users and modularity (Q) of the partition induced by the manual annotation at the party and coalition level.}
    \label{tab:bench}
    \begin{tabularx}{\textwidth}{XXXXX}
    \toprule
    \textbf{Dataset}
    & 
    \textbf{Nodes \rev{in GCC}}
    & 
    \textbf{Links} & 
    \textbf{$Q(\text{parties})$} &
    \textbf{$Q(\text{coalitions})$}\\
    \midrule
    \textbf{President} & 169 & 219 & 0.763 & 0.618 \\
    \textbf{Crisis}    & 145 & 246 & 0.721 & 0.641 \\
    \textbf{Elections} & 154 & 355 & 0.777 & 0.649 \\
    \bottomrule
    \end{tabularx}
\end{table}

To extrapolate this benchmark to the broader network, we propagated community labels from verified users to the rest of the user population using a label propagation algorithm~\citep{Raghavan2007a}.
While these propagated labels are inherently noisier than manual annotations, they provide a principled proxy for large-scale political alignment and enable a systematic comparison of community-detection strategies over the full dataset.

\rev{
\subsection*{Standard community detection on the full retweet network shows limited political alignment}
}

\begin{figure}[htbp!]
    \centering
    \includegraphics[width=\textwidth]{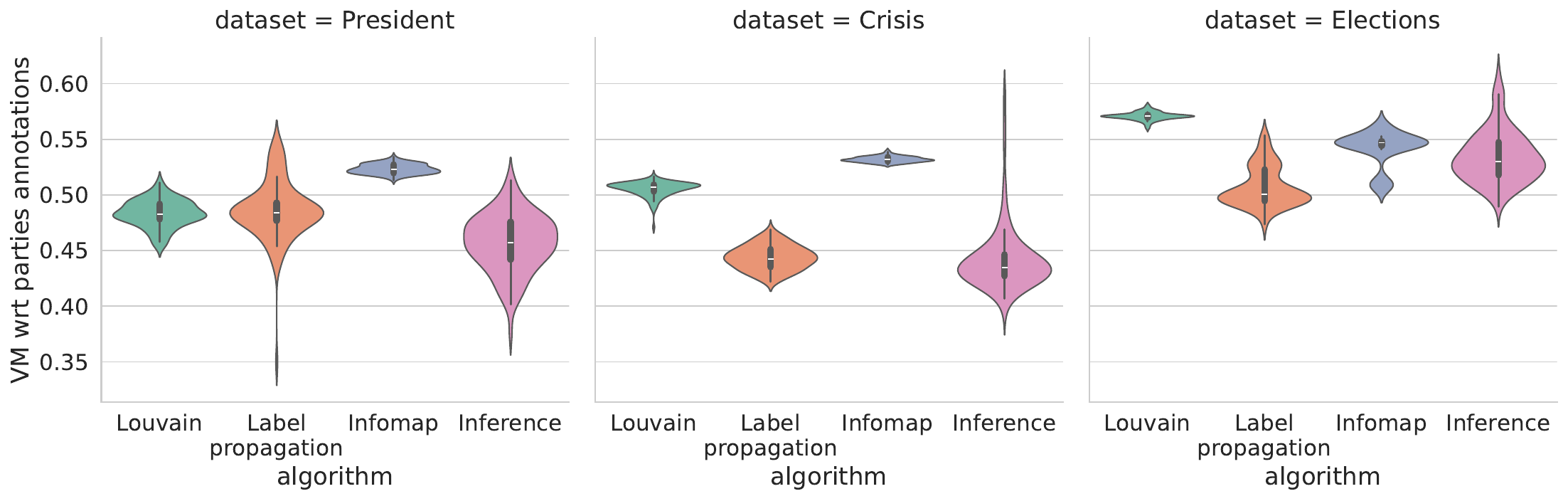}
    \caption{\textbf{Limits of standard methods.} The $\mathrm{VM}_{1}$ between partitions obtained with off-the-shelf algorithms, restricted to the annotated users, and the parties' annotations. Each violin represents the distribution over 100 independent runs. Standard methods show a limited  accuracy in recognizing the affiliation of renowned politicians.}
    \label{fig:stability_accuracy_standard}
\end{figure}

We applied four widely used community detection algorithms—Louvain, label propagation, Infomap, and stochastic block model inference—to the complete retweet networks, \rev{defined as the directed weighted graph including all users active during the event and all observed retweet interactions.}
While all methods identified non-trivial community structure, their agreement with the manual annotations was limited.
V-measure scores for party affiliation ranged between 0.4 and 0.6 (cfr. Fig.~\ref{fig:stability_accuracy_standard}), with substantial variability across algorithm runs.

Comparisons between algorithms further revealed significant inconsistency, with pairwise V-measures for the same dataset lying in the $[0.3,0.7]$ range (cfr. Fig.~\ref{fig:stability_accuracy_standard_vs_Louvain} in the Appendix).
Qualitative inspection confirmed that standard methods often conflated ideologically distinct actors or fragmented coherent political blocs.

\rev{
These results do not indicate a general failure of standard algorithms, but rather highlight the difficulty of recovering politically meaningful communities when they are applied directly to noisy retweet networks without distinguishing between heterogeneous user roles or filtering statistically expected interactions.
}

\rev{
\subsection*{Leader-anchored and statistically validated networks improve community detection}
}

Applying our two proposed methods, $\monodc{V}$ and $\bidc{V}$, resulted in a substantial improvement in community-detection performance.
When verified users were used as anchors within statistically validated networks, both methods produced partitions closely aligned with the annotated political affiliations.

$\monodc{V}$ achieved particularly high accuracy at the party level, reflecting the focused retweet patterns through which politicians tend to amplify content from their own party leadership.
By contrast, $\bidc{V}$ more effectively captured coalition-level structures, as it relies on shared audience behaviour rather than direct interactions among leaders.

\begin{figure}[htbp!]
    \centering
    \includegraphics[width=\textwidth]{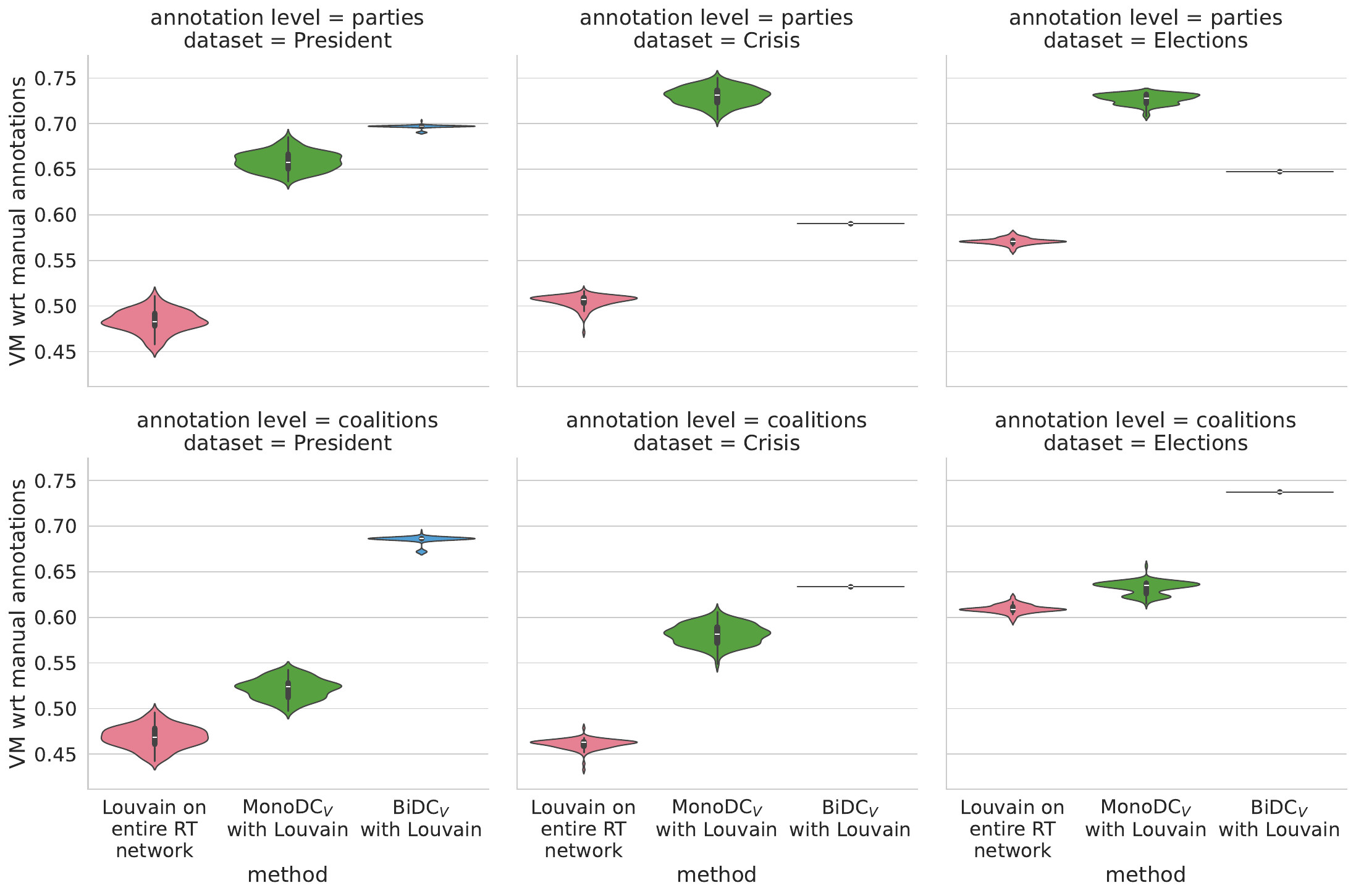}
    \caption{\textbf{Comparison of the performances of $\monodc{V}$ and $\bidc{V}$ with standard methods.} The $\mathrm{VM}_{1}$ between partitions of the annotated users, obtained with different methods, and the annotations at party (first row) and coalition (second row) level. Each violin represents the distribution over 100 independent runs, and we use the Louvain algorithm as a benchmark because it performs no worse than other algorithms (cfr. Fig.~\ref{fig:stability_accuracy_standard}). $\monodc{V}$ and $\bidc{V}$ systematically outperform Louvain, with  $\monodc{V}$ generally preferable in identifying parties while $\bidc{V}$ in identifying coalitions.}
    \label{fig:accuracy_comparison}
\end{figure}

\begin{figure}[htbp!]
    \centering
    \includegraphics[width=\textwidth]{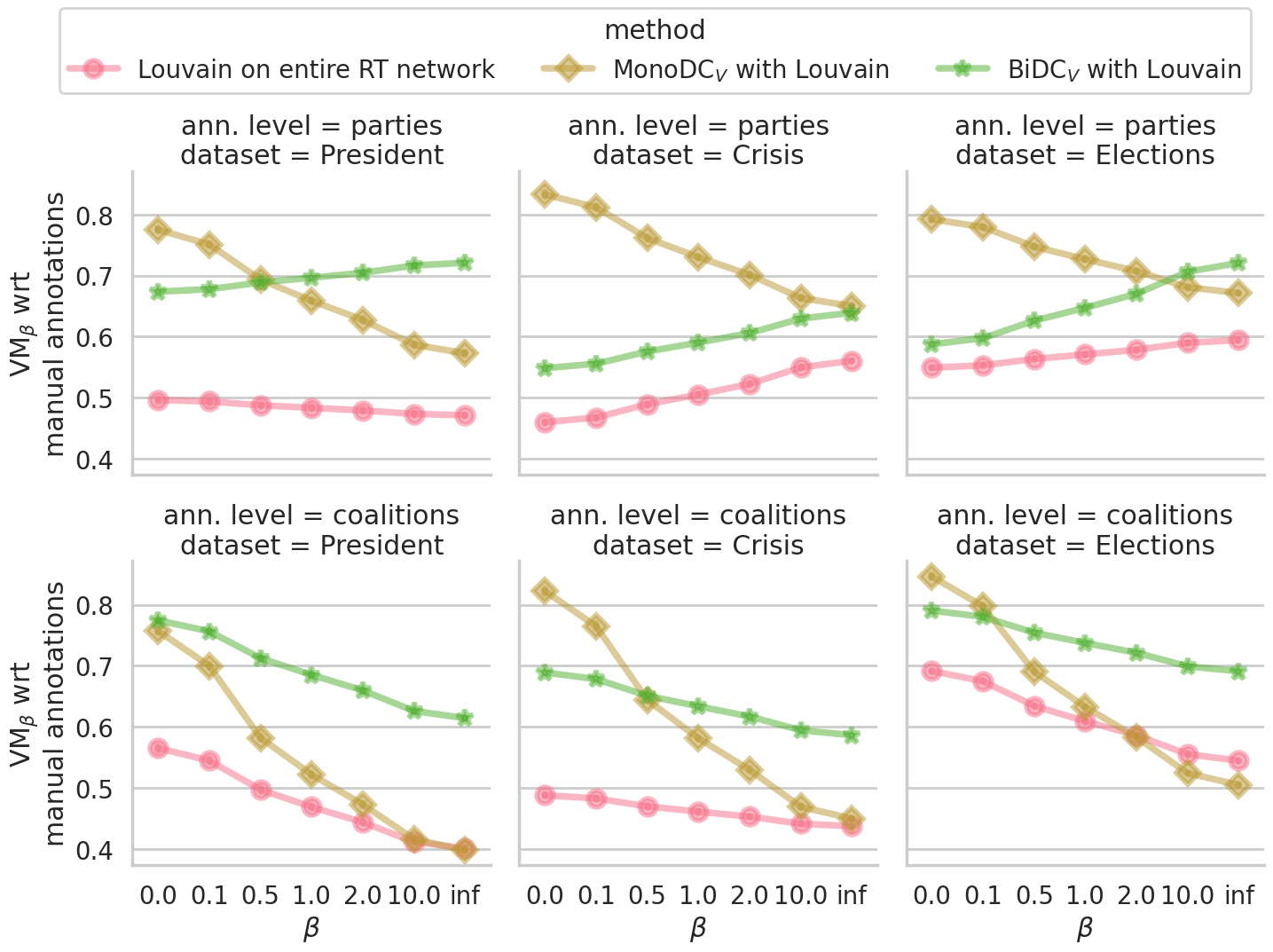}
    
    \caption{\textbf{Comparison of the performances of $\monodc{V}$ and $\bidc{V}$ with standard methods at different resolution levels.} The $\mathrm{VM}_{\beta}$ between partitions of the annotated users, obtained with different methods, and the annotations, for different values of $\beta$ and both at the parties (top) and coalitions (bottom) level.
    Each point is the average over 100 runs of the method indicated by the colour and marker.
    Analogous results are obtained when considering the propagated annotations. When compared with manually annotated parties (top) and coalitions (bottom), $\monodc{V}$ and $\bidc{V}$ exhibit a consistently greater average $\mathrm{VM}_{\beta}$ than standard methods even for $\beta\neq 1$, i.e. at essentially all scales at which the political spectrum can be observed. In this sense, standard algorithms, when executed on the entire retweet network, do not return sub- or super-groups of the political parties that $\monodc{V}$ and $\bidc{V}$ cannot detect.
    }
    \label{fig:VM_beta}
\end{figure}

Quantitatively, V-measure scores for both methods systematically exceeded those of standard algorithms across all datasets.
$\monodc{V}$ reached values above 0.75 for party-level annotations, while $\bidc{V}$ achieved comparably strong results at the coalition level (cfr. Fig.~\ref{fig:accuracy_comparison}).
These findings remained robust across different values of the resolution parameter $\beta$ (cfr. Fig.~\ref{fig:VM_beta}) and when considering propagated labels on the full network (cfr. Fig.~\ref{fig:accuracy_comparison_props} in the Appendix).

\rev{
As a baseline, we apply the Louvain algorithm to the complete retweet network.
This representation includes neither statistical filtering nor a distinction between verified and non-verified users, and thus reflects the raw interaction structure dominated by highly heterogeneous activity patterns.
}

\rev{
Although $\monodc{V}$ and $\bidc{V}$ also rely on Louvain for the final partitioning step, the comparison serves a different purpose.
In our framework, Louvain is applied after maximum-entropy filtering and on networks anchored to a coherent set of political leaders.
Running Louvain directly on the raw retweet network therefore isolates the contribution of preprocessing and leader selection, rather than acting as a competing methodological alternative.
Notably, the Louvain algorithm performs no worse than the others in recovering the annotations (cfr. Fig.~\ref{fig:stability_accuracy_standard}).
}

\rev{
Overall, these results indicate that verified accounts function as effective anchors for political discursive communities \textit{within a statistically validated framework}.
The improved performance does not stem from verified status alone, but from its combination with maximum-entropy filtering and with the institutional coherence of the verified-user set under the pre-2022 verification regime.
}

\rev{
\subsection*{MonoDC captures party structure, while BiDC reveals coalitional organization}
}

The complementary strengths of $\monodc{V}$ and $\bidc{V}$ arise from the nature of their respective network constructions.
$\monodc{V}$ relies on direct retweet interactions among verified users, which tend to encode endorsement and intra-party signalling.
As a result, its statistically validated networks retain a large fraction of edges (94–97\%), indicating dense and coherent communication among political actors.

$\bidc{V}$, by contrast, infers similarity among leaders through audience overlap.
After statistical validation, only 7–12\% of edges are retained, reflecting the fact that only a small subset of shared audiences carries meaningful political alignment.
This filtering allows $\bidc{V}$ to surface stable coalition-level structures, particularly in phases where electoral alliances are clearly defined.

\begin{figure}[htbp!]
    \centering
    \includegraphics[width=.6\textwidth]{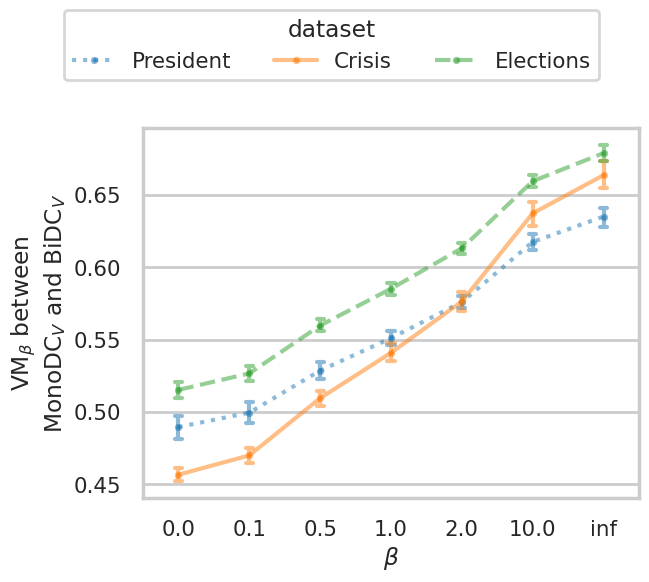}
    \caption{{\textbf{Different resolution of $\monodc{V}$ and $\bidc{V}$.}} 
    The $\mathrm{VM}_{\beta}$, for variable $\beta$, between partitions of all users obtained running  $\monodc{V}$ and  $\bidc{V}$. The clusters found by $\monodc{V}$ are subclusters of those found by $\bidc{V}$, at least in part, and especially for the two datasets, Crisis and Elections, which refer to a political phase in which the coalitions were more clearly defined.
    }\label{fig:mono_vs_bi_beta_ita_elections}
\end{figure}

Consistently, the $\mathrm{VM}_{\beta}$ between the partitions produced by the two methods shows that $\monodc{V}$ communities tend to be subclusters of those identified by $\bidc{V}$, especially for the Crisis and Elections datasets (cfr. Fig.~\ref{fig:mono_vs_bi_beta_ita_elections}).

\begin{figure}[htbp]
\centering
    \begin{subfigure}[b]{.325\textwidth}
        \centering
        \includegraphics[width=\textwidth]{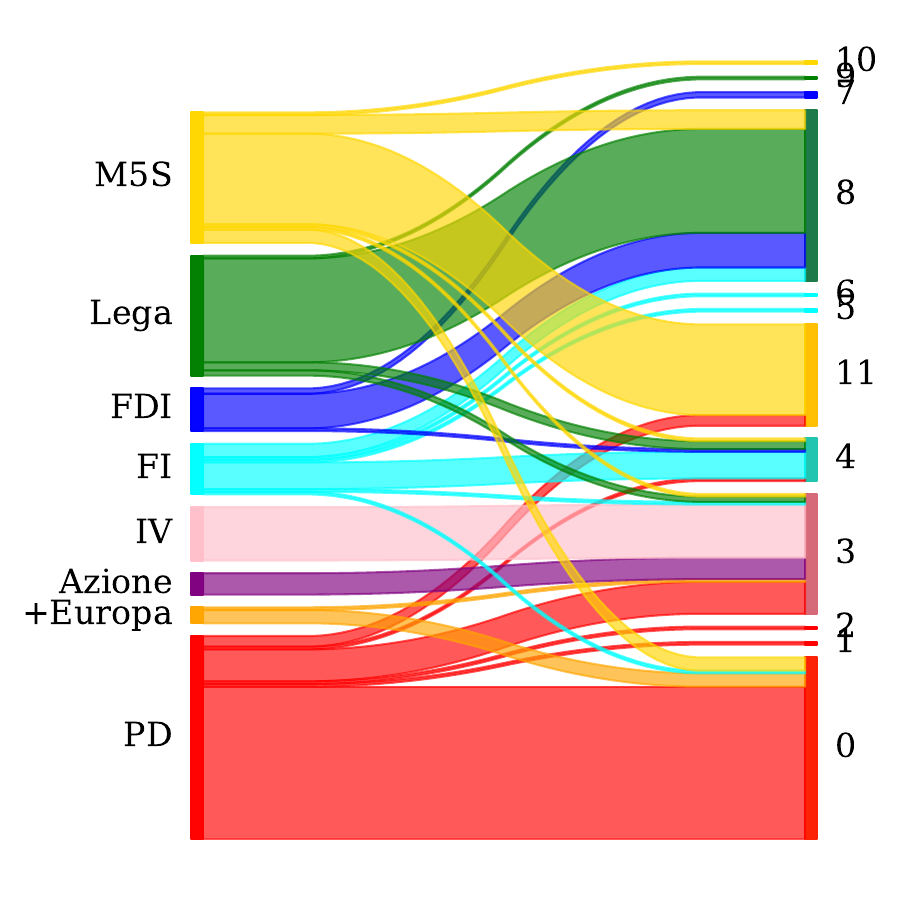}
        \caption{Louvain on the retweet network.}
        \label{fig:sankey_louvain}
    \end{subfigure}
    \begin{subfigure}[b]{.325\textwidth}
        \centering
        \includegraphics[width=\textwidth]{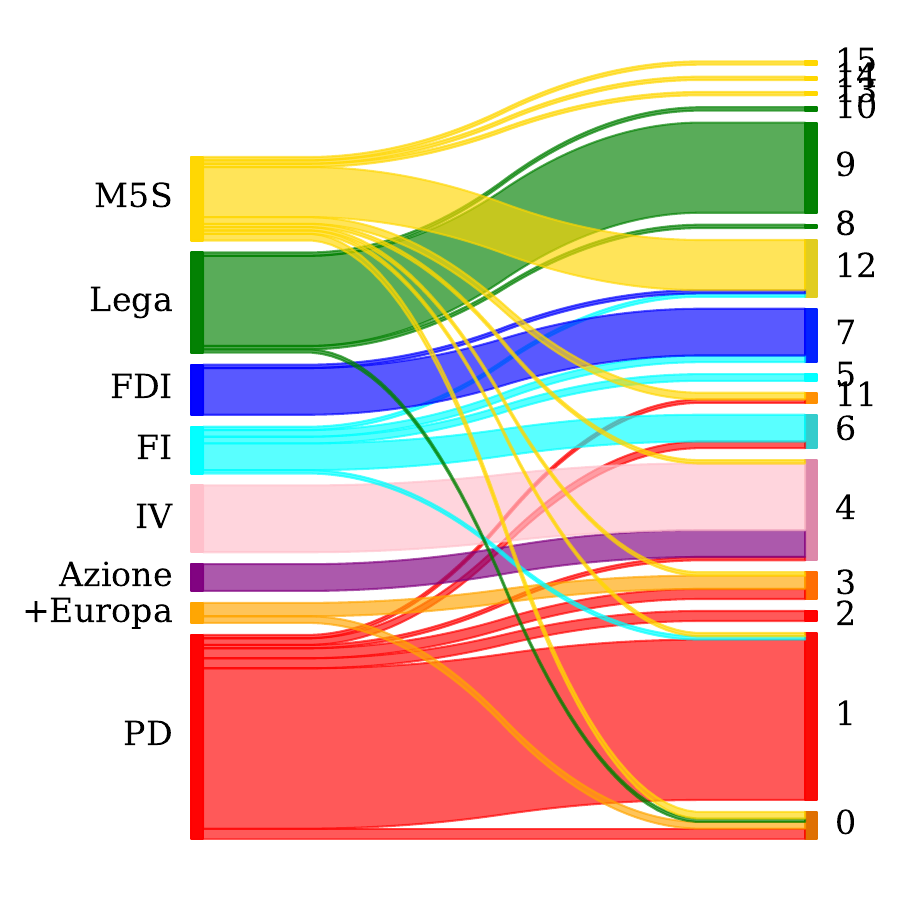}
        \caption{$\monodc{V}$.}
        \label{fig:sankey_mono}
    \end{subfigure}
    \begin{subfigure}[b]{.325\textwidth}
        \centering
        \includegraphics[width=\textwidth]{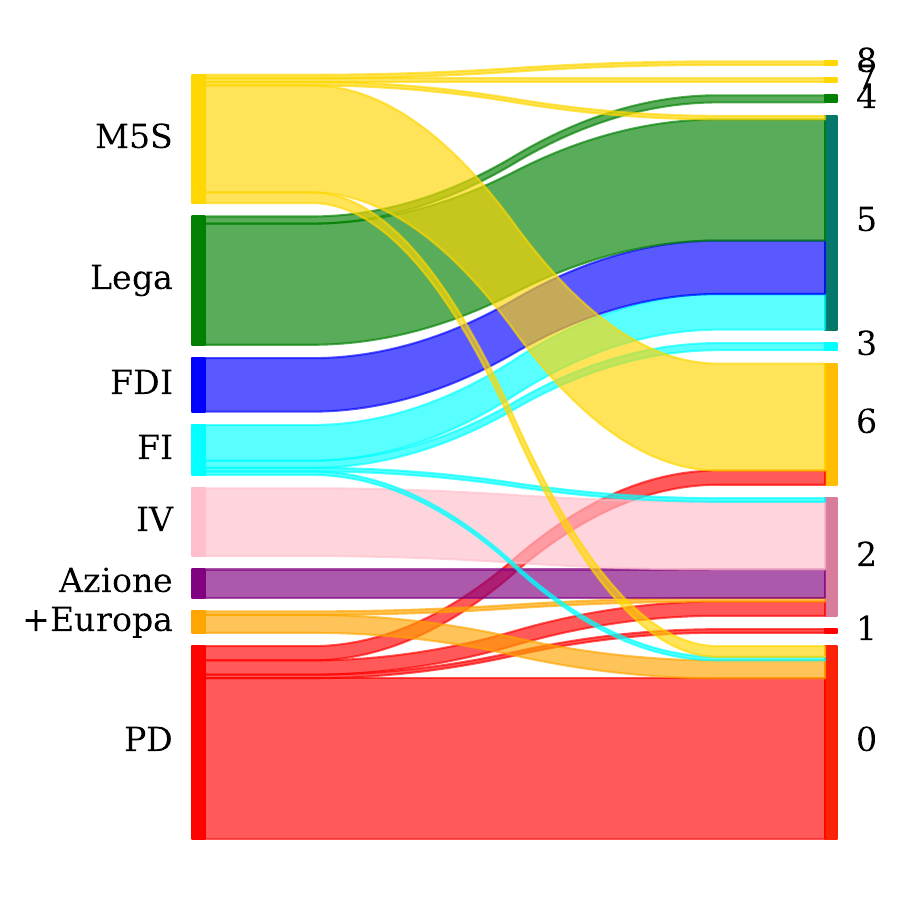}
        \caption{$\bidc{V}$.}
        \label{fig:sankey_bi}
    \end{subfigure}
\caption{\textbf{Qualitative comparison of $\monodc{V}$ and $\bidc{V}$ with Louvain on the Elections dataset.} 
For each of the considered methods, the Sankey diagrams show the composition of the extracted communities in terms of political parties. $\monodc{V}$ very accurately detects parties (Azione and Italia Viva ran together as a single list at the elections), plus some political movements/trends within parties; $\bidc{V}$ almost perfectly detects the official political coalitions of those elections; Louvain, executed on the entire retweet network, provides a more noisy classification that is a mix of the two levels, with Forza Italia separated from its coalitions and many politicians contributing to communities led by some of their rivals.}
\label{fig:sankey}
\end{figure}

A qualitative comparison further illustrates these differences (cfr. Fig.~\ref{fig:sankey}).
$\monodc{V}$ yields fine-grained partitions aligned with individual parties, whereas $\bidc{V}$ aggregates these into broader coalitional structures.
In contrast, Louvain applied to the full retweet network produces a noisier mapping that conflates party- and coalition-level affiliations.

\rev{
We attribute this fragmentation not to political complexity, but to the structural noise of the raw retweet network, which is substantially reduced by maximum-entropy filtering.
}

\rev{
\subsection*{Activity-based definitions of leaders yield weaker political partitions}
}

To assess the robustness of our findings, we replaced verified users with alternative activity-based definitions of the leader set $\top$.
In the case $\top=I$, we selected influential users following~\citep{GonzalezBailon2013}, while $\top=H$ was defined using an adaptation of the Hirsch index~\citep{Hirsch2005} to retweet activity\rev{: a user $u$'s H-index in the debate is the largest $h$ such that at least $h$ tweets authored by $u$ received at least $h$ retweets each, and we define $H$ as the set of users having H-index$\geq3$ (cfr. Fig.~\ref{fig:elbow_H} in the Appendix for a justification of the threshold)}.

\begin{figure}[htbp]
\centering
    \begin{subfigure}[b]{\textwidth}
        \centering
        \includegraphics[width=\textwidth]{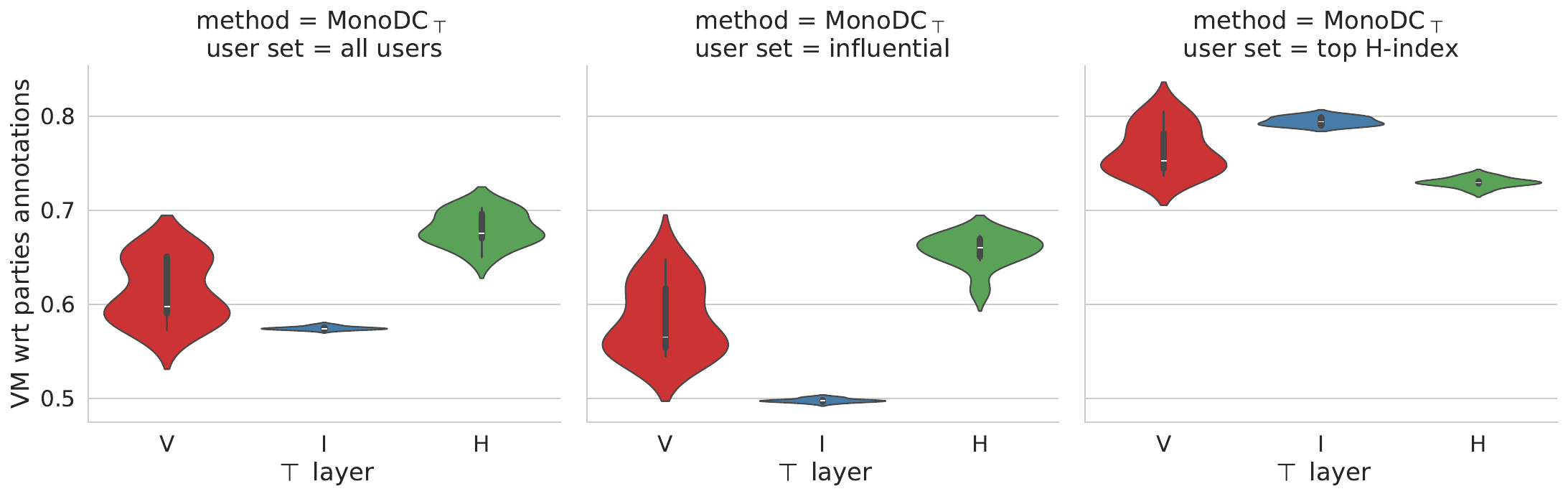}
        \caption{$\monodc{\top}$, for different choices of $\top$, compared with annotations at the parties level.}
        \label{fig:other_mono}
    \end{subfigure}
    \begin{subfigure}[b]{\textwidth}
        \centering
        \includegraphics[width=\textwidth]{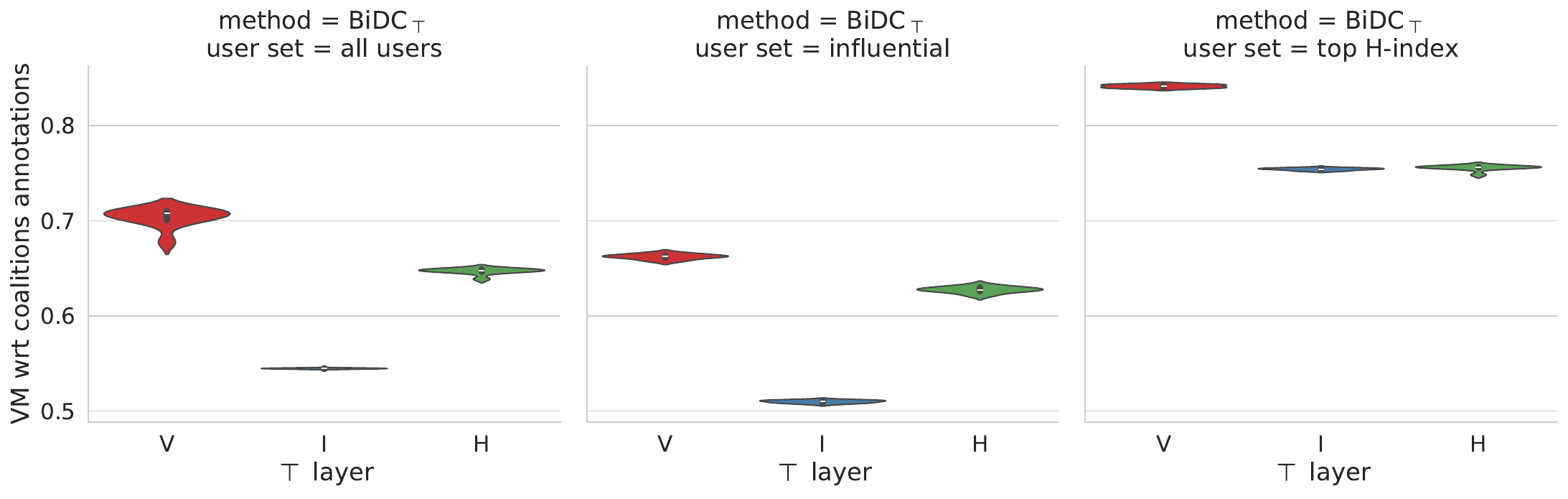}
        \caption{$\bidc{\top}$, for different choices of $\top$, compared with annotations at the coalitions level.}
        \label{fig:other_bi}
    \end{subfigure}
\caption{\textbf{Comparison of the partitions obtained with $\monodc{\top}$ and $\bidc{\top}$, for different choices of $\top$.}
The $\mathrm{VM}_{1}$ between partitions obtained with different choices of $\top$ and the annotations, for the Elections dataset. For each method, we used the level of annotation that the method more accurately reproduces, that is, parties for $\monodc{\top}$ (a) and coalitions for $\bidc{\top}$ (b).
The columns refer to the user set upon which the VM is computed. For instance, ``user set = influential'' means that in the second column we only consider how influential users have been labeled by different methods, temporarily ignoring the rest of the network.
}
\label{fig:biproj}
\end{figure}

In all cases, performance deteriorated relative to anchoring communities on verified users (cfr. Fig.~\ref{fig:biproj}).
While activity-based criteria identify highly active accounts, their discursive behavior is more heterogeneous and less predictive of political alignment.
In particular, H-index-based selections include journalists and commentators whose audiences span ideological boundaries, blurring community structure.

\rev{
These results highlight that the effectiveness of our framework depends not only on selecting active users, but on identifying content creators whose publicly recognized roles constrain and stabilize their online behavior. \rev{Further comparison can be found in Section~\ref{app:comparisons} of the Appendix.}
}

\section*{Discussion}

Digital platforms play an increasingly influential role in shaping the structure and visibility of public discourse.
One crucial contribution of this work is the characterization of a general methodology for extracting meaningful discursive communities from noisy social data.
\rev{
We introduced a leader-based framework that explicitly accounts for the asymmetric structure of online participation.
By separating content producers from their audiences, our approach allows communities to emerge more clearly than methods that treat all users symmetrically.
We showed that identifying and positioning content producers is a necessary first step, after which the broader user base can be reliably assigned.
}

Both proposed methods--$\monodc{}$ and $\bidc{}$--filter user interactions through maximum entropy null models, thereby revealing latent organizational patterns otherwise obscured by platform dynamics or behavioral heterogeneity.
\rev{
In the specific case considered here, we applied this framework to datasets centered on highly politicized debates, where the expected discursive structures correspond to political parties and electoral coalitions.
Within this context, the distinction between the two approaches offers analytical flexibility: $\monodc{}$ emphasizes direct political signaling through retweets, while $\bidc{}$ captures broader audience-based alignments, including coalition-level structures.
}

\rev{In addition, through a systematic comparison of three alternative criteria for selecting leading users,} our findings demonstrate that institutional status, rather than raw activity, provides the strongest signal for organizing online debates.
\rev{
We found, in fact, that pre-2022 verified accounts represent the most informative anchors for community detection.
We interpret this result as a consequence of their public recognizability and the behavioral constraints imposed by their off-platform roles, rather than as an intrinsic property of verification itself. 

While the institutional meaning of verification has changed following the introduction of paid subscriptions, the methodological insights of this study remain valid.
In particular, our results suggest that carefully curated sets of publicly relevant actors--identified either through institutional roles or external information--can serve as effective seeds for analyzing online discourse, at least in the well-studied and highly structured context of political debates.
This insight opens the door to more efficient data-collection strategies, in which the ego networks of a limited number of high-relevance users are sufficient to recover much of the political structure of online debates, even under restrictive platform access conditions. Indeed, similar strategies have been implemented, for instance in~\cite{Brugnoli2024}.
More broadly, our study reinforces the idea that the organization of digital public debate potentially depends not only on individual user behavior, but also on how platforms encode and signal authority.
While the methods proposed here are technical, they contribute to a growing body of work showing that discursive structure, legitimacy, and visibility are deeply intertwined in online public spheres.
}

\rev{
As any other product of research, our analyses have some limitations.
First, our empirical results pertain exclusively to the pre-2022 verification regime, under which verified accounts corresponded to publicly recognized figures and formed coherent, politically aligned communities.
In this historical setting, verification acted as a reliable proxy for public relevance and institutional embeddedness, providing a stable anchor for discursive structures.
}
\rev{
The transition to paid verification introduces a qualitatively different regime, whose effects fall outside the scope of the present analysis.
Concerns about potential distortions of online discourse--such as increased visibility for actors lacking public accountability or the erosion of previously coherent community structures--should therefore be interpreted as speculative and grounded in plausible mechanisms rather than in direct empirical evidence.
Whether political discourse under the new regime will remain similarly structured, fragment into less interpretable patterns, or reorganize around alternative signals of authority remains an open empirical question.
}
Future research should explicitly address this gap by analyzing data collected after the introduction of paid verification, enabling direct comparisons between the two regimes.
Such studies could assess whether the structural role once played by verified users has weakened, persisted, or been replaced by other indicators of public relevance.

\rev{Second, our empirical analysis focuses on political debates within a single national context, Italy, and on a single social media platform, Twitter/X.
This choice reflects both data availability and methodological considerations: analyzing a political system that is well known to the authors allows for more reliable manual annotations and a more accurate interpretation of political actors, coalitions, and discursive dynamics.
Importantly, this focus does not represent an intrinsic limitation of the proposed framework, which is grounded in structural features of online discourse--such as strong participation asymmetries and the presence of discussion leaders--that have been widely documented across platforms, languages, and cultural settings~\cite{Watts2007,Hilbert2017,Bracciale2018}.
Consistently with this perspective, previous studies have already shown that $\bidc{}$ performs well in different countries~\cite{Mattei2022,Declerck2022a,DeClerck2024}, while the present work provides the first systematic investigation of $\monodc{}$ and a direct comparison between alternative leader-based constructions.
Moreover, the proposed approach does not rely on platform-specific features, but rather on general interaction patterns between content producers and their audiences, making its application to other platforms--where visibility and influence may be structured through different mechanisms--a natural direction for future research.}

\section*{Competing interests}
  The authors declare that they have no competing interests.

\section*{Author's contributions}
S.G. and F.S. conceived and designed the study. F.S. and A.M. collected the data. S.G., F.S. and A.M. performed the modeling and data analysis with input from G.C. S.G and A.M. contributed to data preprocessing and visualization. S.G. and F.S. interpreted the results and drafted the manuscript. All authors contributed to revisions and approved the final version of the manuscript.

\section*{Acknowledgements}
  The authors are thankful to Lorenzo Federico for interesting discussions.

\section*{Funding}
  S.G. and F.S. were partially supported by the project ``CODE – Coupling Opinion Dynamics with Epidemics'', funded under PNRR Mission 4 ``Education and Research'' - Component C2 - Investment 1.1 - Next Generation EU ``Fund for National Research Program and Projects of Significant National Interest'' PRIN 2022 PNRR, grant code P2022AKRZ9.

\bibliography{bmc_article} 
\newpage
\appendix 

\section{Maximum Entropy Null Models}\label{app:maxent}

We start from a real network $G^*$ (in the following, all quantities related to the observed network will be denoted by an asterisk $*$). Firstly, let us define a set $\mathcal{G}$ -- the {\em ensemble}-- including all possible graphs having the same number of nodes $N$ as in $G^*$\footnote{Since the number of nodes is fixed in the randomization, we drop the asterisk $*$ from $N$.}. Let us identify a vector of quantities $\vec{C}$ that are fundamental for the description of the real system. A proper benchmark for $G^*$ would be completely random, but for the information contained in $\vec{C}(G^*)$: in this way it is possible to highlight all properties of $G^*$ that are not trivially due to $\vec{C}$. Translating this {\em rationale} in maths, we need to perform a constrained maximization of the Shannon entropy (to have a maximally random benchmark) associated to the ensemble, where the constraints are exactly $\vec{C}(G^*)$. In this sense, the (ensemble) average $\langle\vec{C}\rangle$ will be set exactly to $\vec{C}(G^*)$.

If $S$ is the Shannon entropy, defined as 
\begin{equation*}
    S=-\sum_{G\in\mathcal{G}}P(G)\ln P(G), 
\end{equation*}
then we can perform the constrained maximisation using the method of Lagrange multipliers, i.e. by maximising $S'$ defined as 

\begin{equation*}
    S'=S+\vec{\theta}\cdot\Big(\vec{C}(G^*)-\langle\vec{C}\rangle\Big)+\alpha\Big(1-\sum_{G\in\mathcal{G}}P(G)\Big),
\end{equation*}
where $\vec{\theta}$ is the vector of Lagrange multipliers associated with $\vec{C}$ and $\alpha$ is the Lagrangian multiplier associated to the normalization of the probability $P(G)$. The probability solving the maximization problem above has the standard functional form of exponential random graphs:

\begin{equation}\label{eq:ERG_p}
    P(G)=\frac{e^{-\vec{\theta}\cdot\vec{C}(G)}}{Z(\vec{\theta})},
\end{equation}
where $Z=e^{\alpha+1}=\sum_{G\in\mathcal{G}}e^{-\vec{\theta}\cdot\vec{C}(G)}$ is the partition function. In order to get the numerical value of the Lagrangian multipliers $\vec{\theta}$ we have to explicitly impose 
\begin{equation}\label{eq:maxlikelihood}
    \langle \vec{C}\rangle=\vec{C}(G^*).
\end{equation}
It can be shown that the condition in Eq.~\eqref{eq:maxlikelihood} can also be obtained through the maximization of the (log-)likelihood~\citep{Garlaschelli2008}.

The framework described above is particularly flexible and can be adapted to different kinds of networks. In the following, we are interested in its bipartite versions for bipartite undirected and bipartite directed networks, i.e. respectively, the Bipartite Configuration Model (or \emph{BiCM}~\citep{Saracco2015}) and the Bipartite Directed Configuration Model (or \emph{BiDCM}~\citep{vanLidth2019,Becatti2019d}). In both cases, the constraints imposed in the entropy maximisation are, respectively, the undirected and directed degree sequences.

\subsection{Bipartite Configuration Model (BiCM)}

Given a bipartite network with layers $\top$ (e.g., users) and $\perp$ (e.g., tweets), let $N_\top$ and $N_\bot$ be their dimensions.
Nodes belonging to the $\top-$layer will be denoted by Latin indices, while the ones belonging to the $\bot-$layer will be denoted by Greek ones. Any bipartite network is completely described by its biadjacency matrix $\mathbf{B}$, i.e. a $N_\top\times N_\bot$ matrix whose entries $b_{i\alpha}=1$ if nodes $i\in\top$ is connected to $\alpha\in\bot$ and $b_{i\alpha}=0$ otherwise. The degree of a node $i\in\top$ ($\alpha\in\bot$), i.e. the number of connections that the nodes have, is simply $k_i=\sum_\alpha b_{i\alpha}$ ($h_\alpha=\sum_i b_{i\alpha}$).

The Bipartite Configuration Model preserves the degree sequences $k_i = \sum_\alpha b_{i\alpha}$ and $h_\alpha = \sum_i b_{i\alpha}$.
Following the procedure sketched above, the probability of observing a graph $G_{\text{Bi}}$ in the ensemble is:
\[
P(G_{\text{Bi}}) = \prod_{i,\alpha} \frac{e^{-(\theta_i + \eta_\alpha) b_{i\alpha}(G_{\text{Bi}})}}{1 + e^{-(\theta_i + \eta_\alpha)}},
\]
where $\vec{\theta}$ and $\vec{\eta}$ are the Lagrangian multipliers associated with the degree sequences on, respectively, layers $\top$ and $\bot$.

The corresponding link probabilities are:
\[
p_{i\alpha} = \frac{e^{-(\theta_i + \eta_\alpha)}}{1 + e^{-(\theta_i + \eta_\alpha)}}.
\]
To get the numerical values of the Lagrangian multipliers, we can impose the likelihood maximization~\citep{Garlaschelli2008}.

\subsubsection{Validated Projection under BiCM}

Using BiCM as a benchmark permits to validate the co-occurrences in the real network~\citep{Saracco2017}. In fact, observed co-occurrences are simply the ``projection'' of the information contained in the bipartite network into one of the two layers, say $\top$:
\begin{equation*}
    V^{ij}=\sum_\alpha b_{i\alpha}b_{j\alpha},\,\forall i\neq j\in\top
\end{equation*}
(in Ref.~\citep{Saracco2015} co-occurrences were also called \emph{V-motifs} since, if the two layers of the bipartite networks are drawn horizontally, they sketch a ``V'' between the layers).
Since in BiCM the probabilities per link are independent, then 
\begin{equation*}
    \langle V^{ij}\rangle=\sum_\alpha p_{i\alpha}p_{j\alpha},\,\forall i\neq j\in\top.
\end{equation*}
In practice, the entire BiCM distribution of each $V^{ij}$ can be calculated: they are Poisson-Binomial distributions, i.e. the extension of Binomial distribution in which each event has a different probability. Calculating exactly the Poisson-Binomial distribution is computationally costly. It was shown, that when probabilities are relatively small --as in our case--, the Poisson-Binomial distribution can be finely approximated with a Poisson distribution~\citep{Hong2013}. Therefore, it is possible to efficently calculate the p-value associated with each observed co-occurrence $(V^{ij})^*$. Since the number of different co-occurrences is $\binom{N_\top}{2}$, we need multiple hypothesis testing. In the present manuscript, we implemented the False Discovery Rate (\emph{FDR}, ~\citep{Benjamini1995}) since it controls the rate of False Positives. Again, also the validated projection used in the present manuscript was calculated using the python module \href{https://pypi.org/project/bicm/}{\texttt{bicm}}.

\subsection{Bipartite Directed Configuration Model (BiDCM)}

Let us now consider the case in which we have a bipartite directed network and we want our model to include the information about the in- and out-degrees of nodes on both layers. It can be shown that the probability per bipartite directed graph $G_\text{BiD}$ factorises in terms of the BiCM probability per two undirected bipartite graphs $G_\text{Bi}^{\top\rightarrow\bot}$ and $G_\text{Bi}^{\top\leftarrow\bot}$, describing, respectively, all edges from  layer $\top$ to layer $\bot$, and vice versa~\citep{vanLidth2019,Becatti2019d}:
\begin{equation*}
    P_\text{BiDCM}(G_\text{BiD})=P_\text{BiCM}(G_\text{Bi}^{\top\rightarrow\bot})\cdot P_\text{BiCM}(G_\text{Bi}^{\top\leftarrow\bot}). 
\end{equation*}
If $b_{i\alpha}^{\top\rightarrow\bot}$ and $b_{i\alpha}^{\top\leftarrow\bot}$ are the generic entries of the biadjacency matrices associated to, respectively, $G_\text{Bi}^{\top\rightarrow\bot}$ and $G_\text{Bi}^{\top\leftarrow\bot}$, then:
\begin{equation*}
\begin{split}
    P_\text{BiDCM}(G_\text{BiD})=&\Big[\prod_{i,\alpha}(q_{i\alpha}^{\top\rightarrow\bot})^{b_{i\alpha}^{\top\rightarrow\bot}}(1-q_{i\alpha}^{\top\rightarrow\bot})^{(1-b_{i\alpha}^{\top\rightarrow\bot}})\Big]\\
        &\cdot\Big[\prod_{i',\alpha'}(q_{i'\alpha'}^{\top\leftarrow\bot})^{b_{i\alpha}^{\top\leftarrow\bot}}(1-q_{i\alpha}^{\top\leftarrow\bot})^{(1-b_{i\alpha}^{\top\leftarrow\bot}})\Big],
\end{split}
\end{equation*}
where 
$q_{i\alpha}^{\top\rightarrow\bot}$ and $q_{i\alpha}^{\top\leftarrow\bot}$ are the (independent) probability to observe a link from $i$ to $\alpha$ or vice versa. Following the same procedure described above for BiCM, $q_{i\alpha}$s can be expressed in terms of the Lagrangian multipliers associated with the (constrained) maximization of the entropy, i.e. 
\begin{equation*}
    \begin{split}
    q_{i\alpha}^{\top\rightarrow\bot}=&\dfrac{e^{-(\theta_i^\text{out}+\eta_\alpha^\text{in})}}{1+e^{-(\theta_i^\text{out}+\eta_\alpha^\text{in})}};\\
    q_{i\alpha}^{\top\leftarrow\bot}=&\dfrac{e^{-(\theta_i^\text{in}+\eta_\alpha^\text{out})}}{1+e^{-(\theta_i^\text{in}+\eta_\alpha^\text{out})}}.
    \end{split}
\end{equation*}
The numerical values of the $\theta_i^\text{out}$, $\theta_i^\text{in}$, $\eta_\alpha^\text{out}$ and $\eta_\alpha^\text{in}$ can be obtained through the maximization of the likelihood~\citep{Garlaschelli2008}. %The validation procedure induced by BiDCM of the directed co-occurrence network can be found in Appendix~\ref{app:bidcm_validation}.

\subsubsection{Validated Projection under BiDCM}\label{app:bidcm_validation}

As done for the BiCM, we can provide a statistical validation of the monopartite multilink directed network obtained by projecting, say, on layer $\top$, a bipartite directed network. The number of edges from node $i$ to node $j$ in the projection is 
\begin{equation*}
    \mathcal{V}^{ij}=\sum_\alpha b_{i\alpha}^{\top\rightarrow\bot} b_{j\alpha}^{\top\leftarrow\bot}.
\end{equation*}
The probability distribution of $\mathcal{V}^{ij}$, following the BiDCM, is again a Poisson-Binomial and, since also in the present case the probabilities are small, the Poisson-Binomial can be finely approximated by a Poisson distribution. Therefore we are interested in calculating $\langle\mathcal{V}^{ij}\rangle$ because it is the only parameter of the Poisson distribution. Since all $b_{i\alpha}^{\top\rightarrow\bot}$ and $b_{i\alpha}^{\top\leftarrow\bot}$ are independent, it is simply
\begin{equation}\label{eq:v_bidcm}
    \langle\mathcal{V}^{ij}\rangle=\sum_\alpha q_{i\alpha}^{\top\rightarrow\bot} q_{j\alpha}^{\top\leftarrow\bot}.
\end{equation}

\subsubsection{BiDCM-Validated Projection for Users and Posts}
In the analyses above, the BiDCM-validated projection was used to validate the multilink retweet network among users. The original system is a bipartite directed network in which the two layers $\top$ and $\bot$ are, respectively, users and tweets: an arrow goes from a user to a tweet if the user is the author of the tweet and vice versa if the user retweeted the tweet. Since every post has a single author, in $G_\text{Bi}^{\top\rightarrow\bot}$ each node $\alpha\in\bot$ has $\kappa_\alpha^\text{in}=\sum_i b_{i\alpha}^{\top\rightarrow\bot}=1$. With such an identification, the probability that the user $i$ is the author of the post $\alpha$ is independent of $\alpha$ and reads:
\begin{equation*}
    q_{i\alpha}^{\top\rightarrow\bot}=\dfrac{e^{-(\theta_i^\text{out}+\eta^\text{in})}}{1+e^{-(\theta_i^\text{out}+\eta^\text{in})}};
\end{equation*}
in this case, the Bipartite Configuration Model reduces to a Bipartite Partial Configuration Model~\citep{Saracco2017}.
Imposing that the constraints, i.e. the number of tweets authored by each user and the fact that each post has a single author, returns
\begin{equation}\label{eq:bidcm_trick}
    q_{i\alpha}^{\top\rightarrow\bot}=\dfrac{\kappa_i^\text{out}}{N_\bot},\quad\forall\alpha\in\bot.
\end{equation}
Such identification is particularly effective in our case, since 
plugging Eq.~\eqref{eq:bidcm_trick} in Eq.~\eqref{eq:v_bidcm}, gives
\begin{equation}\label{eq:v_bidcm_trick}
    \langle\mathcal{V}^{ij}\rangle=\sum_\alpha q_{i\alpha}^{\top\rightarrow\bot} q_{j\alpha}^{\top\leftarrow\bot}=\dfrac{\kappa_i^\text{out}}{N_\bot}\sum_\alpha q_{j\alpha}^{\top\leftarrow\bot}=\dfrac{\kappa_i^\text{out}\kappa_j^\text{in}}{N_\bot}, 
\end{equation}
where we used that the degree sequence is constrained in $G_\text{Bi}^{\top\leftarrow\bot}$. The result in Eq.~\eqref{eq:v_bidcm_trick} is remarkable: the expected value $\langle\mathcal{V}^{ij}\rangle$ can be easily calculated without knowing explicitly the value of the Lagrangian multipliers. Therefore, p-values against BiDCM of the observed $\mathcal{V}^{ij}$ can be easily calculated, and further statistically validated using the FDR procedure described above~\citep{Benjamini1995}. 

\subsubsection{Limitations}
Besides the simplicity of Eq.~\eqref{eq:v_bidcm_trick}, there is an important issue, already mentioned for the present application in~\citep{Becatti2019d}. In the null model implemented above, constraints are satisfied on average. Therefore, there are configurations in the ensemble that are not physical, i.e. posts that may have either more than a single author or none. Since both non-physical configurations, i.e. more authors or none, contribute to calculating the p-values in opposite directions, their final effect is hard to predict. %\fab{\textbf{[Aggiungiamo qualche conto sulla frequenza di posts senza autore/con piu' di un autore nel ensemble?]}} 
Nevertheless, in light of the present results and the ones of previous research, the agreement with the ground truth seems to indicate that, at least in the present application, the final effect of such an approximation is limited. Such an issue is going to be the target of further research.

\subsection{Implementation}

In the paper, we used the implementation of the BiCM and BiDCM available in the python module \href{https://pypi.org/project/NEMtropy/}{\texttt{NEMtropy}}, described in~\citep{Vallarano2021}.
\href{https://pypi.org/project/bicm/}{\texttt{bicm}} is also available as a standalone module.
As a reference, the running time for BiDCM on the largest dataset are reported in Table~\ref{tab:calculation_time_data}.

\begin{table}[htbp]
    \centering
    \caption{\textbf{Mean and standard deviation of the execution time, expressed in seconds, of the different steps of $\bidc{V}$ on the Elections dataset, the largest dataset considered in this paper.} These timings highlight that the cost of $\bidc{V}$ is dominated by the label propagation algorithm presented in~\citep{Raghavan2007a}.}
    \label{tab:calculation_time_data}
\begin{tabular}{lrr}
\toprule
 & mean & std \\
algorithm step &  &  \\
\midrule
Label propagation on rt network & 943.76 & 62.14 \\
Louvain on top network & 2.64 & 0.03 \\
validated projection p=0.01 & 69.74 & 2.42 \\
\bottomrule
\end{tabular}
\end{table}

\section{V-Measure for Clustering Similarity}\label{app:v_measure}

To assess the similarity between two ways of partitioning the same set of users, we use of the V-measure ($\mathrm{VM}_{\beta}$), a parametric measure of cluster similarity proposed in~\citep{rosenberg2007v}.

Let $S(P)$ be the Shannon entropy of $P$, and $S(P\mid Q)$ be the conditional entropy of $P$ given $Q$.
If $P$ and $Q$ are two partitions, the $\mathrm{VM}_{\beta}(P,Q)$ is defined as:
\[
\mathrm{VM}_{\beta}(P,Q)=\frac{(1+\beta)I(P,Q)}{S(P)+\beta S(Q)}
\]
where $I(P,Q)$ is the mutual information between $P$ and $Q$.
$VM_\beta$ can be interpreted as the harmonic mean, weighted by $\beta$, of homogeneity $h$ and completeness $c$:
\[
h(P, Q) = 1 - \frac{S(P | Q)}{S(P)}, \quad
c(P, Q) = 1 - \frac{S(Q | P)}{S(Q)},
\]
\[
VM_\beta(P, Q) = \frac{(1 + \beta) h c}{\beta h + c}.
\]
Note that $h=1$ if and only if the clusters obtained with $Q$ are sub-clusters of those obtained with $P$, whereas $c=1$ if and only if the clusters obtained with $Q$ are super-clusters of those obtained with $P$~\citep{rosenberg2007v}.

For $\beta=1$ we have 
\[
\mathrm{VM}_{1}(P,Q) = \frac{2I(P,Q)}{S(P)+S(Q)} = 1-\frac{VI(P,Q)}{S(P)+S(Q)}
\] 
i.e., the $\mathrm{VM}_{1}$ accounts for the mutual information between the two partitions in a way that does not depend on their entropy, being equivalent to the Variation of Information~\citep{Meila2007}, normalized by the total entropy of the partitions.
With respect to other information-theoretic metrics, the $\mathrm{VM}_{1}$ and the VI have the advantage of being two distances -- in particular, of satisfying the triangle inequality.
Further, the $\mathrm{VM}_{1}$ always takes values in $[0,1]$, making it possible to directly compare the scores obtained for different combinations of algorithms and/or datasets. 
By letting $\beta$ vary in $[0,+\infty)$, we consider both scenarios in which one partition consists of sub-communities (e.g., political movements/trends within parties) or super-communities (e.g., political coalitions) of the other one.

\begin{figure}[htbp!]
    \centering
    \includegraphics[width=\textwidth]{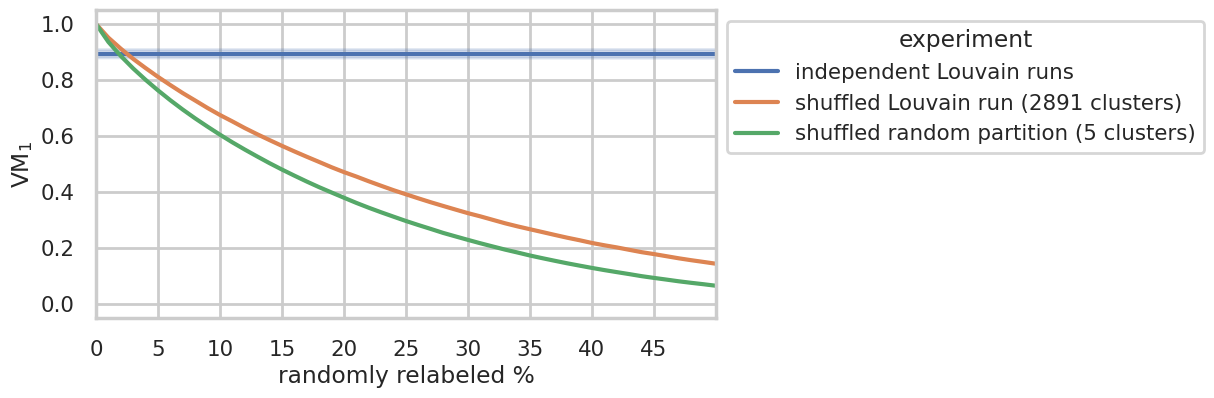}
    \caption{\textbf{Understanding the values of $\mathrm{VM}_1$.} The plot shows $\mathrm{VM}_1$ between independent runs of the Louvain algorithm compared with the $\mathrm{VM}_1$ between some partition of the Elections dataset --either 5 equally sized random communities, or the communities obtained running the Louvain algorithm-- and a noisy version of the same partition obtained randomly reallocating part of the users, as a function of the percentage of the network that is reallocated.}
    \label{fig:VM_benchmark}
\end{figure}

It is worth noting that the VM of two partitions does not decrease linearly with the number of elements on which the two partitions disagree.
Fig.~\ref{fig:VM_benchmark} shows the average $\mathrm{VM}_1$ between some partition of the Elections dataset and a noisy version of the same partition obtained randomly relabeling part of the users.
We consider two choices for the initial labeling: 5 equally sized random communities and the communities obtained running the Louvain algorithm.
In both cases, the $\mathrm{VM}_1$ rapidly decreases to about 0.8 when $\sim$5\% of the network has been relabeled, and to 0.6 when $\sim$10-13\% of the network has been relabeled.
We also show that two different runs of the Louvain algorithm generate partitions whose $\mathrm{VM}_1$ is about 0.9, on average. 
Based on these tests, we deduce that a reasonably good $\mathrm{VM}_1$ between two partitions obtained with different methods/algorithms lies in the range $[0.6-0.9]$.

\rev{
\section{Data collection and political context}\label{app:data_collection}
Data were downloaded using Twitter/X official APIs, using a Researcher account. Data were collected using the Python module \texttt{twarc2} (\url{https://twarc-project.readthedocs.io/en/latest/twarc2_en_us/}), using the \texttt{search --archive} command to access the archive of data. 

\begin{table}[htbp]
    \centering
    \caption{Description of the queries used for the data collection}
    \label{tab:data}
    \begin{tabularx}{\textwidth}{XXXX}
    \toprule
    \textbf{Dataset} & 
    \textbf{Keywords} &
    \textbf{Start time} &
    \textbf{End time} \\
    \midrule
    %%%%%%%%%%%%%%%%%%%%%%%%%
    \textbf{President}&%\tnote{1} &
    quirinale &
    2022-01-01 &
    2022-01-31\\
    & presidente della repubblica &&\\
    & mattarella &&\\
    %%%%%%%%%%%%%%%%%%%%%%%%%
    \midrule
    \textbf{Crisis}%\tnote{2}
    & crisidigoverno
    & 2022-07-01
    & 2022-08-01
    \\
    & crisigoverno &&\\
    & draghi &&\\
    & governo &&\\
    %%%%%%%%%%%%%%%%%%%%%%%%%
    \midrule
    \textbf{Elections}&%\tnote{3} &
    elezionipolitiche2022 & 2022-08-25 &
    2022-10-02\\
    & elezionipolitiche22 &&\\
    & elezioni &&\\
    & PD &&\\
    & Letta &&\\
    & Calenda &&\\ 
    & Azione &&\\
    & Renzi &&\\
    & ItaliaViva &&\\ 
    & FdI &&\\ 
    & Meloni &&\\ 
    & Lega &&\\ 
    & Salvini &&\\ 
    & Conte &&\\ 
    & M5S &&\\ 
    & ForzaItalia &&\\ 
    & Berlusconi &&\\
    \bottomrule
    \end{tabularx}
\end{table}
In the President dataset, ``Quirinale'' is the name of the Roman hill where the headquarters of the Italian President of the Republic is. ``Presidente della Repubblica'' is the Italian name for President of the Republic, while ``Mattarella'' is the surname of the president that was confirmed in early 2022.\\ 
In 2021, President Mattarella consistently declined to make himself available for a second term, despite sustained pressure from a broad range of political actors. Nonetheless, in the absence of an officially endorsed candidacy, he continued to attract a significant number of votes in successive ballots, obtaining 336 votes in the sixth round and 387 in the seventh. By 29 January, the failure of all alternative candidacies and the persistent coordination breakdown between the two major coalitions rendered the re-election of the incumbent a focal point of convergence. Following this deadlock, Mattarella accepted a renewed mandate and was re-elected with 759 votes.\\

In Italian political discourse, ``crisi di governo'' is an idiomatic expression referring to a cabinet crisis, namely the fall of the executive without a systemic constitutional breakdown. Mario Draghi was the Prime Minister at the time of the crisis.\\ 
On July 13, 2022, during a press conference, Conte, leader of the M5S,  declared that his party would abstain from voting on the Aid Decree, which the government had designated as a matter of confidence. The following day, the Senate approved the decree, while M5S representatives withdrew from the chamber during the vote. Although this action did not constitute a formal withdrawal of governmental support, it was broadly interpreted as a substantial challenge to the administration's agenda and effectively precipitated a political crisis within Draghi's Cabinet. After strong debates in the two chambers of the Italian parliament, on July 21, Draghi definitively submitted his resignation to President Mattarella, who was compelled to accept, as the government's loss of parliamentary support had by then become unequivocal and irreversible.\\

The keywords used to gather tweets for the Elections dataset include the surname of the main political party leaders (i.e. Letta, Calenda, Renzi, Meloni, Salvini, Conte, Berlusconi), as well as the name of their party (PD, Azione, Italia Viva, Fratelli d'Italia, Lega, Movimento 5 Stelle, Forza Italia); other keywords relate to the event itself.\\ 
Following the Draghi government's collapse and resulting parliamentary deadlock, President Sergio Mattarella dissolved Parliament on July 21, 2022, calling early elections for September 25. The centre-right coalition led by Giorgia Meloni's national-conservative Fratelli d'Italia (FdI) won an absolute parliamentary majority. Meloni was appointed Prime Minister on October 22, becoming Italy's first female head of government.
}

\section{Experimental Details and Extended Comparisons}\label{app:comparisons}

\subsection{Community Detection Algorithms and Implementations}

Community detection algorithms used:
\begin{itemize}
    \item \textbf{Louvain, Label Propagation, Infomap} from the \texttt{igraph} Python package.
    \item \textbf{Stochastic Block Model (SBM)} inference via \texttt{graph-tool}'s \texttt{minimize\_blockmodel\_dl()}.
\end{itemize}
The \texttt{minimize\_blockmodel\_dl()} function fits a stochastic block-model to the given graph by minimizing the model's description length.
To guide the inference process, we set it explicitly to search for a planted-partition, that is, exclusively for assortative communities.
In addition, we refined the algorithm as recommended in graph-tool's documentation, that is, by running 100 times the \texttt{multiflip\_mcmc\_sweep()} method with zero temperature and 100 iterations.

We ran each algorithm 100 times. Each single run of the Louvain algorithm is already obtained taking the partition with the greatest modularity out of 100 independent executions with shuffled vertices. Similarly, each run of the Label propagation algorithm is already obtained by taking, for each node, the most frequent label out of 100 independent executions.

\subsection{Definition of $H$ Users}

We repeated MonoDC and BiDC using:
\begin{itemize}
    \item \textbf{Influential users} ($I$): users with follower/followee ratio $>1$ and more mentions received than made.
    \item \textbf{Top H-index users} ($H$): users with $H \geq 3$ where $H$ is the largest number such that the user has at least $H$ tweets each with at least $H$ retweets.
\end{itemize}

\begin{figure}[htbp]
    \centering
    \includegraphics[width=.6\textwidth]{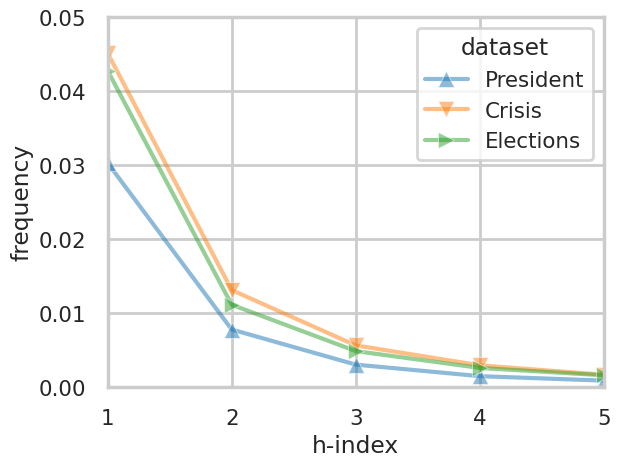}
    \caption{\textbf{The distribution of the H-index in the three datasets.} The threshold $H=3$ used to tell apart $\top$ and $\bot$ users is justified by the empirical H-index distribution having an elbow at 3; cumulatively, only about $1-2\%$ of the users pass the threshold.}
    \label{fig:elbow_H}
\end{figure}

The threshold $H>3$ has been selected based on the elbow-method applied to the frequency of users for whom $H>h$ for different values of $h$ (cfr. Fig.~\ref{fig:elbow_H}).

\subsection{Effect of Significance Thresholds}

\begin{figure}[htbp!]
    \centering
    \includegraphics[width=\textwidth]{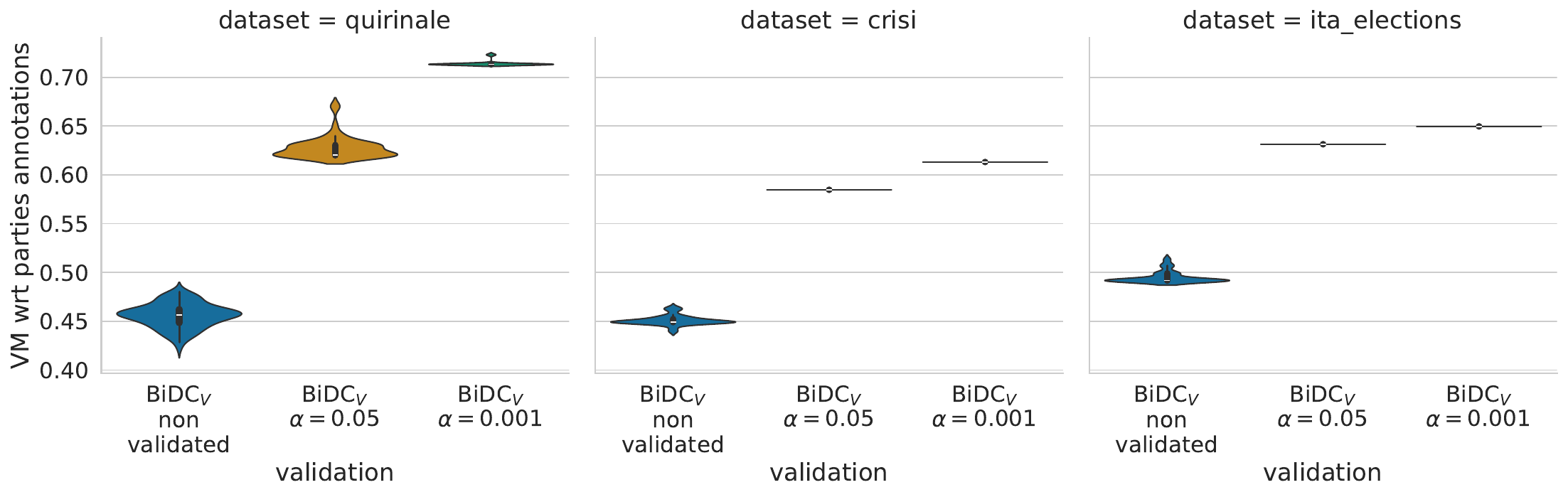}
    \includegraphics[width=\textwidth]{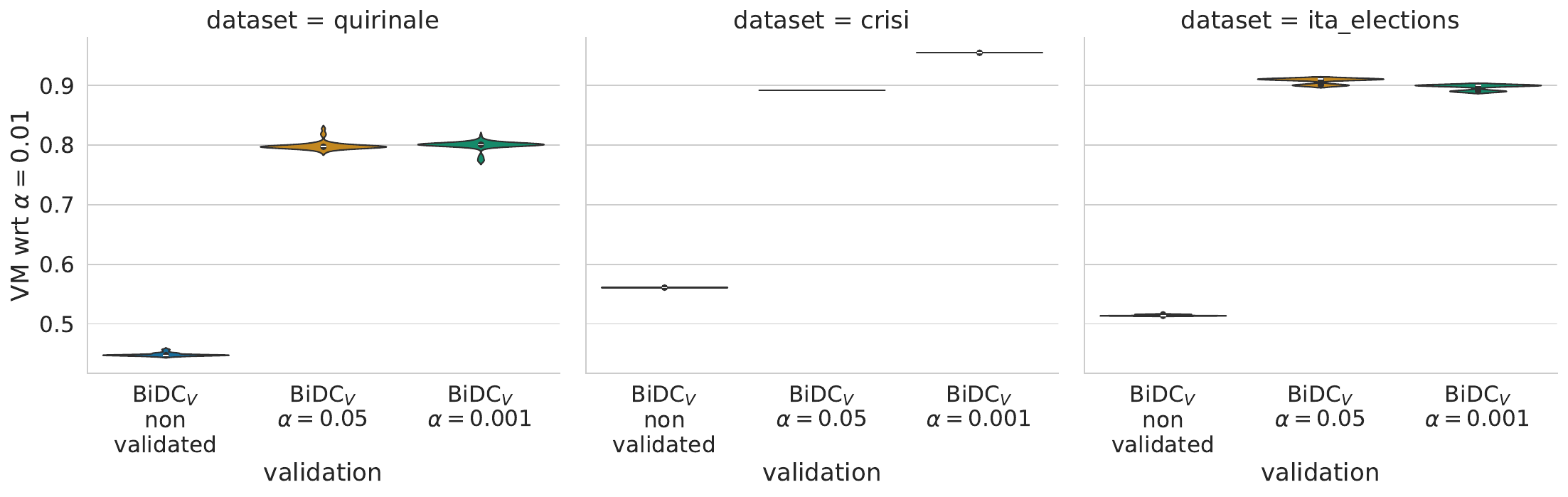}
    \caption{\textbf{An analysis of the role of validation ($\alpha$ denotes the significance level) in determining the partitions obtained for the $\top$ layer of $\bidc{V}$.} Top: the $\mathrm{VM}_1$ between partitions of annotated $V$ users, obtained for different validation levels, and the annotations. Bottom: the $\mathrm{VM}_1$ between partitions of $V$ users, obtained for different validation levels, and those obtained with $\alpha=0.01$. Each point is a different run of the method indicated by the color and marker.
    The results on the non-validated network are very poor, whereas the significance level has a limited impact on the inferred partitions.}
    \label{fig:validation_levels}
\end{figure}

When running the $\bidc{V}$ method, we observed that only a small percentage of all edges resulted as being statistically significant.
We therefore explored BiDCM validation at multiple significance levels ($\alpha = 0.001$, $0.01$, $0.05$). Results, shown in Fig.~\ref{fig:validation_levels}, were stable across values, confirming robustness. Non-validated projections performed poorly.

\subsection{Additional Comparisons}

\begin{figure}[htbp!]
    \centering
    \includegraphics[width=\textwidth]{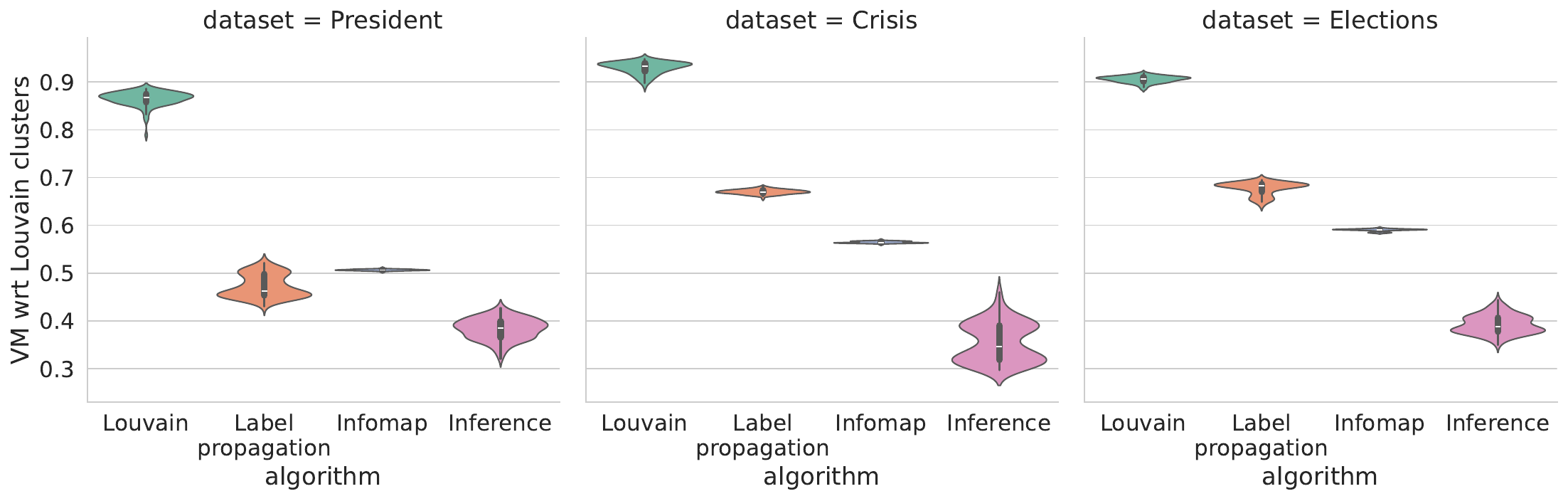}
    \caption{\textbf{The average $\mathrm{VM}_{1}$ between partitions obtained with off-the-shelf algorithms and 100 runs of the Louvain algorithm (in green, the self-similarity of Louvain partitions).} Louvain was chosen as the benchmark, but all standard methods display comparable performances. Standard algorithms show great variability in the communities assigned to verified users.}
    \label{fig:stability_accuracy_standard_vs_Louvain}
\end{figure}

\begin{figure}[htbp!]
    \centering
    \includegraphics[width=\textwidth]{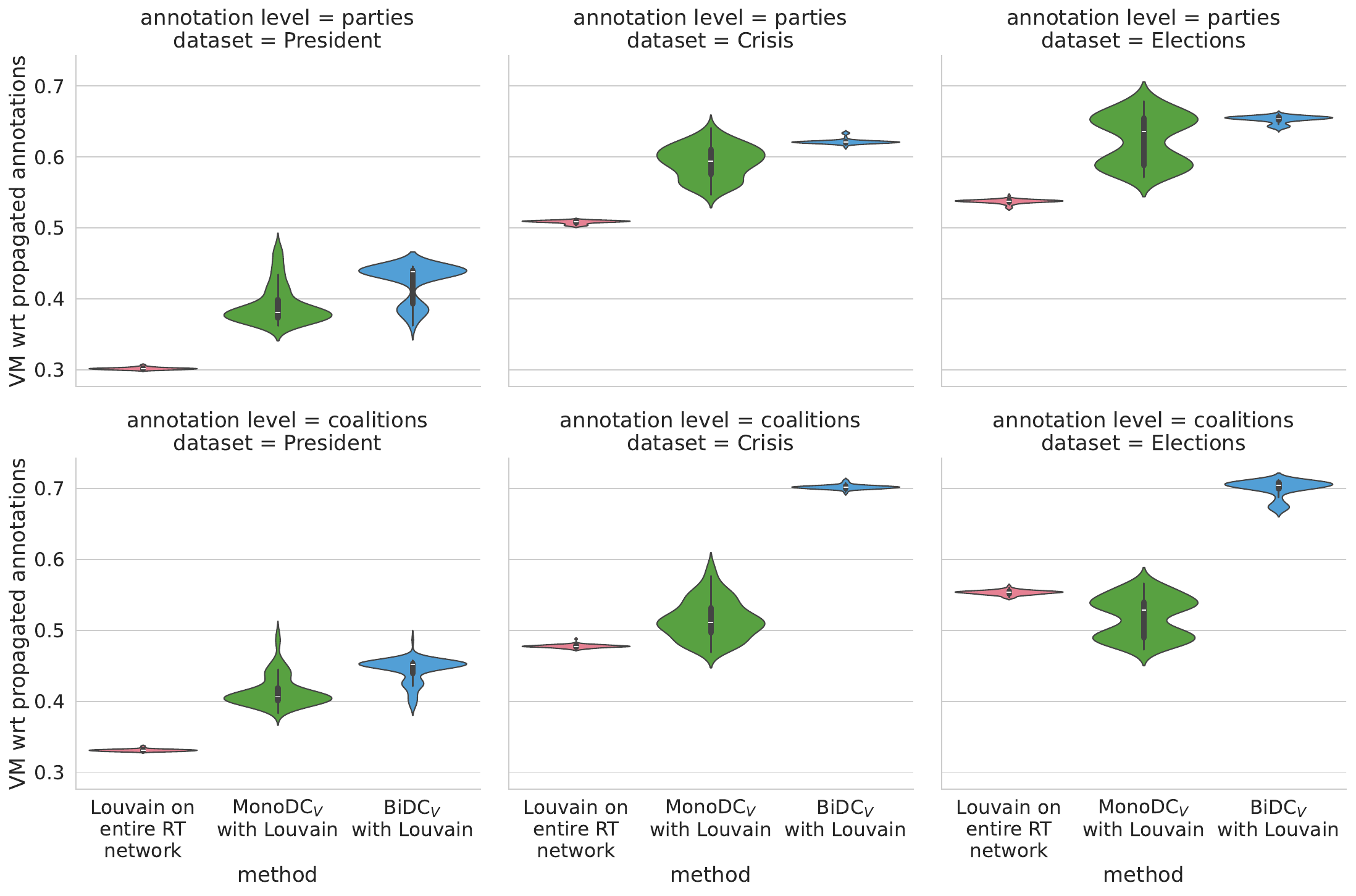}
    \caption{\textbf{The $\mathrm{VM}_{1}$ between partitions of the entire network, obtained with different methods, and the propagated annotations.} Let us note that, when comparing the results of the propagated labels of $\monodc{V}$ and $\bidc{}$ with the annotations of verified users propagated using the same label propagation algorithm, our methods have an advantage. Nonetheless, the performances of standard methods are worse, even when limited to annotated users.\\
}
    % Each point is a different run of the method indicated by the color and marker.}
    \label{fig:accuracy_comparison_props}
\end{figure}

\begin{figure}[htbp]
\centering
    \includegraphics[width=\textwidth]{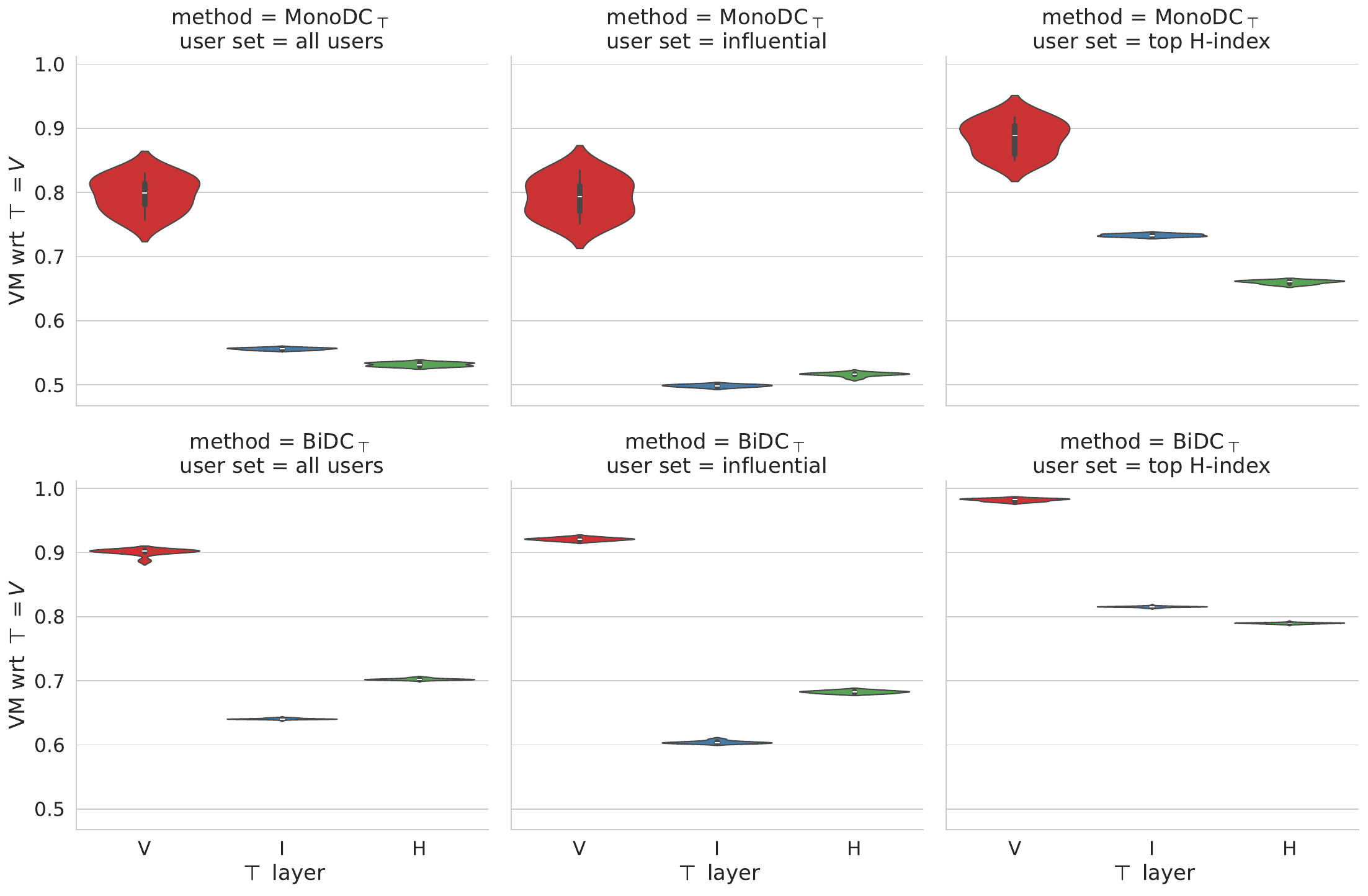}
    \caption{\textbf{Cross-comparison of the partitions obtained with $\monodc{\top}$ and $\bidc{\top}$, for different choices of $\top$.}
The $\mathrm{VM}_{1}$ between partitions obtained with different choices of $\top$ and those obtained with $\top=V$, used as a reference, for the Elections dataset. 
The columns refer to the user set upon which the VM is computed. For instance, ``user set = influential'' means that in the second column we only consider how influential users have been labeled by different methods, temporarily ignoring the rest of the network.
% Each point is a different run of the method indicated by the color and marker on the Elections dataset.
}
\label{fig:biproj_cross}
\end{figure}

As mentioned in the main article, we performed the following additional tests that overall confirm the main findings:

\begin{itemize}
    \item A comparison of the partitions obtained with different off-the-shelf clustering algorithms, using the Louvain method as a benchmark (cfr. Fig.~\ref{fig:stability_accuracy_standard_vs_Louvain}).
    \item A comparison of the partitions obtained with different methods for the entire network and the propagated annotations (cfr. Fig.~\ref{fig:accuracy_comparison_props}).
    \item A cross-comparison of the partitions obtained with $\monodc{\top}$ and $\bidc{\top}$, for different choices of $\top$ (cfr. Fig.~\ref{fig:biproj_cross}).
\end{itemize}

\end{document}